\documentclass[preprint,aps,floatfix]{revtex4}
\usepackage{dcolumn}%
\usepackage{amssymb}%

\begin{document}


\def\dee{{\cal D}}
\def\prod{{\cal M}}
\def \INFZ{{\rm Inf}[(\gamma/Z)_{\rm L},\nu]}
\def \SSQW{\sin^2\theta_W}
\def \SSQWSQ{\sin^4\theta_W}
\def \a{{(\uparrow)}}
\def \b{{(\downarrow)}}
\def\bz{\beta_{{}_Z}}
\def \ZHtrans{$Z\HIGGS$-transverse}
\def \ZHlong{$Z\HIGGS$-longitudinal}
\def \BETA{\bz}
\def \GAMMA{\gamma_{{}_Z}}
\def \GINV{\sqrt{1{-}\BETA^2}}
\def \ts{\thinspace}
\def \nts{\negthinspace}
\def \longitudinal{longitudinal}
\def \amp{{\cal A}}
\def \Ihat{\widehat{\cal I}}
\def \thetas{\theta^{*}}
\def \T{{\cal S}}
\def \G{{\cal G}}
\def \eebar{e^{+}e^{-}}
\def \WWbar{W^{+}W^{-}}
\def \ginv{\gamma_{{}_Z}^{-1}}
\def \GeV{{\rm \enspace GeV}}
\def \beq{\begin{equation}}
\def \eeq{\end{equation}}
\def \beqa{\begin{eqnarray}}
\def \eeqa{\end{eqnarray}}
\def\cphi{c_\varphi}
\def\sph{s_\varphi}
\def\cph{c_\varphi}
\def \cxi{c_\xi}
\def \sxi{s_\xi}
\def \cchi{c_\chi}
\def \schi{s_\chi}
\def \cth{c_\theta}
\def \sth{s_\theta}
\def\DEE{\mathfrak{D}}
\def\ECKS{{\cal Z}}
\def\Em{{\cal M}}
\def\half{\hbox{$1\over2$}}
%
%
%
%
%
%
\def\HIGGS{h}


\preprint{
  \parbox{2in}{Fermilab--Pub--06/160-T \\
  hep-ph/0606052
}  }

\title{Using Spin Correlations to Distinguish $Zh$ from $ZA$ 
at the International Linear Collider}
\author{Gregory Mahlon}%
\email{gdm10@psu.edu}
\affiliation{Penn State Mont Alto \\
1 Campus Drive, Mont Alto, PA 17237 \\ 
USA }
\author{Stephen Parke}%
\email{parke@fnal.gov}
\affiliation{Department of Theoretical Physics\\
Fermi National Accelerator Laboratory \\
P.O. Box 500, Batavia, IL 60510 \\
USA }
\date{June 5, 2006}
\begin{abstract}
We investigate how to exploit the spin information
imparted to the $Z$ boson in associated Higgs production
at a future linear collider as an aid in distinguishing
between $CP$-even and $CP$-odd Higgs bosons.  
We apply a generalized spin-basis analysis which allows
us to study the possibilities offered by non-traditional
choices of spin projection axis.  In particular, we find
that the $Z$ bosons produced in association with a $CP$-even
Higgs via polarized 
collisions are in a single
transverse spin-state ($>90\%$ purity)
when we use the $Z\HIGGS$-transverse basis, provided that
the $Z$~bosons are not ultra-relativistic (speed $<0.9c$).
This same basis applied to the associated production of a $CP$-odd 
Higgs yields $Z$'s that are an approximately equal mixture of
longitudinal and transverse polarizations.
We present a decay angular distribution 
which could be used to distinguish between the
$CP$-even and $CP$-odd cases.
Finally, we make a few brief remarks about how this distribution 
would be affected if the Higgs boson turns out to not be a
$CP$-eigenstate.
\end{abstract}
\maketitle


\vfill\eject\section{Introduction}

One of the goals of high energy physics during the next decade
is to elucidate the mechanism for the spontaneous symmetry breaking
in the electroweak sector.  In the Standard Model, this symmetry
breaking leaves behind a single Higgs 
boson with spin-parity-charge-conjugation quantum numbers
${\cal J}^{PC} = 0^{++}$.  As soon as a Higgs candidate 
is discovered, we will want to examine it closely to see if its
properties match those of the Standard Model, or, if not, which
extension to or replacement of the Standard Model is implied.
For example, supersymmetric theories contain a $0^{+-}$ state
alongside of a SM-like $0^{++}$ boson;
certain other models add a $Z'$ boson (spin-1) to the mix.
Thus, the direct determination of the ${\cal J}^{PC}$ assignment
for a newly-discovered boson will be a priority.  
There are at least different three ways of constraining
the quantum numbers of such a boson.  First, we can examine
the energy-dependence of the associated production
cross section just above threshold~\cite{Miller}.  
Second, we can look at the dependence of the
cross section on the production angle~\cite{Barger}.
Thirdly, and this is the point of this paper, we
can study a suitably-chosen decay-angle distribution.
Even though this last method is potentially the most difficult, 
it will be useful 
to supplement the other two methods
with additional information from angular correlations to
resolve possible ambiguities 
or to provide a cross-check~\cite{Miller2}.

\begin{figure*}
\vspace*{5cm}
\includegraphics{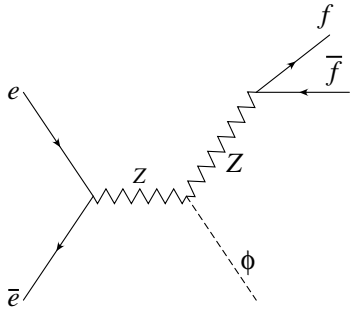}
\vspace{0.0cm}

\caption[]{The Feynman diagram for the production
and decay of a Higgs boson ($\phi$) in association with a $Z$ boson.
}
\label{FeynDiagram}
\end{figure*}

At first glance, it might seem that since the Higgs ($\phi$)
is a spin-0 object,
all angular correlations should be trivial.  However, this is not
the case.   For example, the decay products of the tau leptons
coming from $\phi\rightarrow\tau^{+}\tau^{-}$ decays exhibit 
non-trivial correlations that can be used to probe the nature
of the Higgs~\cite{tautau}.  Similarly,
if the decay to a pair of vector bosons exists, then by looking at
the angular distributions of the vector boson
decay products it is possible
to distinguish a $CP$-even Higgs boson ($\HIGGS$) from 
a $CP$-odd Higgs boson ($A$)~\cite{Barger,moreWWZZ}.
These angular distributions are sensitive to the type of Higgs boson
involved because of the presence of a $\phi VV$ vertex in the process.
However, this method suffers from the drawback that there is no
guarantee that the $\phi \rightarrow VV$ branching ratio will
be large enough to be useful.
Fortunately, it is not necessary
that $\phi \rightarrow VV$ exist to probe the properties of the
Higgs boson:  the so-called associated production mechanism
also contains a $\phi VV$ vertex
(see Fig.~\ref{FeynDiagram}).

Thus, the $Z$ boson produced in association
with the Higgs at a linear collider carries information about the
type of Higgs it was produced with.  Consequently, by examining
the spin state of the $Z$, we can learn about the Higgs in a 
model-independent fashion.  Since we understand $Z$
decays very well (see, for example, the review of $Z$
physics contained in Ref.~\cite{PDB04}),
any deviations from Standard Model predictions will point to 
new physics in the Higgs sector.

Traditionally, angular correlations have been studied within the 
context provided by the helicity basis.  For a light
Higgs and a linear collider
running at a full TeV or so, this is appropriate.
However, in situations
where the $Z$ boson and Higgs are not ultra-relativistic, the
helicity basis may not give the most useful description of the
physics involved.  Instead, it is fruitful to explore other
choices of spin axis~\cite{ttbar1,ttbar2,WHspin,xibasis,WWZZZH}.  
One framework which facilitates this exploration in a
fairly straightforward manner is the generic spin basis ($\xi$-basis)
introduced by Parke and Shadmi in Ref.~\cite{xibasis}.
One possibility that this framework allows for is the analysis of
the data (or independent subsets thereof) in two (or more) different 
ways, to see if the spin content of the $Z$ bosons 
varies with $\xi$ in the predicted manner.

Spin correlations in associated Higgs production at a hadron 
collider  ($q\bar{q}'\rightarrow W\phi$)
have already been studied in Ref.~\cite{WHspin}.
Angular correlations are most easily observed
and understood in the zero momentum frame of the event.
At a hadron collider, however, the
$z$ component of the total momentum in the event is ambiguous,
making the zero momentum frame difficult to find.  
We have written this paper from the point-of-view of 
a future $e^{+}e^{-}$ linear collider to utilize the advantages
offered by such a machine.  Not only is the zero momentum
frame relatively well-known in this case, but the ability to
polarize the beams enhances the angular correlations.

The outline of this paper is as follows.  After a brief discussion
of our notation and conventions in Sec.~\ref{XIbasis}, we present
the polarized $Z\HIGGS$ and $ZA$ production cross sections at a linear
collider in Sec.~\ref{AssocProd}.
Next, we turn to a review of the decay 
distributions ($Z\rightarrow f\bar{f})$
in the case of polarized $Z$ bosons in Sec.~\ref{polZdec}.
In Sec.~\ref{ProdDec} we combine the production and decay
amplitudes to derive expressions for the triply-differential
cross sections for 
$e^{+}e^{-} \rightarrow Zh \rightarrow f\bar{f}h$ and
$e^{+}e^{-} \rightarrow ZA \rightarrow f\bar{f}A$,
with an emphasis on the forms of these distributions in
the helicity and $Z\HIGGS$-transverse bases.
Integrating the triply-differential cross section over
the production and decay azimuthal angles leads to the
principle results of this paper in Sec.~\ref{enchiladas}: 
the angular distribution $d\sigma/d(\cos\chi)$
of the $Z$ boson decay products as seen in the $Z$ rest frame.
We compare this distribution in the $Z\HIGGS$ and $ZA$ cases
in Sec.~\ref{Zh-vs-ZA} and in Fig.~\ref{TheWholeEnchilada}.  
We make a few brief remarks about how this distribution would
be different in the case where the Higgs boson is not
in a $CP$-eigenstate in Sec.~\ref{CPviol}.
Finally, we
end with a summary of our conclusions in Sec.~\ref{CONC}.


\vfill\eject\section{Notation and Conventions}\label{XIbasis}

Throughout this paper,
we use the symbol $\phi$ to refer to a
generic Higgs boson which could either be $CP$ even or $CP$ odd
or have no unique $CP$ eigenvalue.
The symbol $\HIGGS$ will be reserved for use when we are talking
specifically about a $CP$-even Higgs boson.  Finally, a $CP$-odd
Higgs boson will be represented by the symbol~$A$.


\begin{figure*}

\vspace*{13.5cm}
\includegraphics{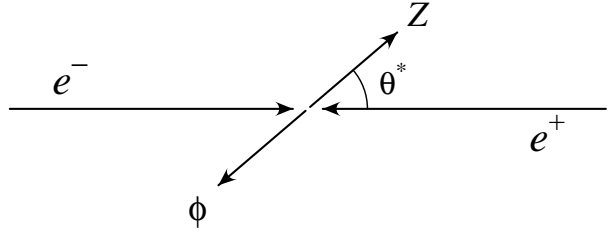}
\includegraphics{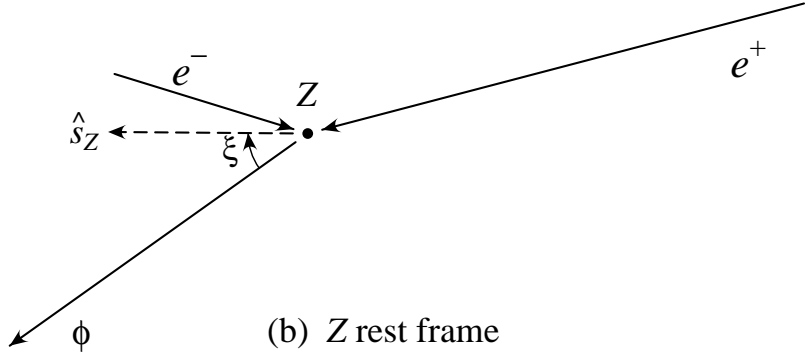}
\includegraphics{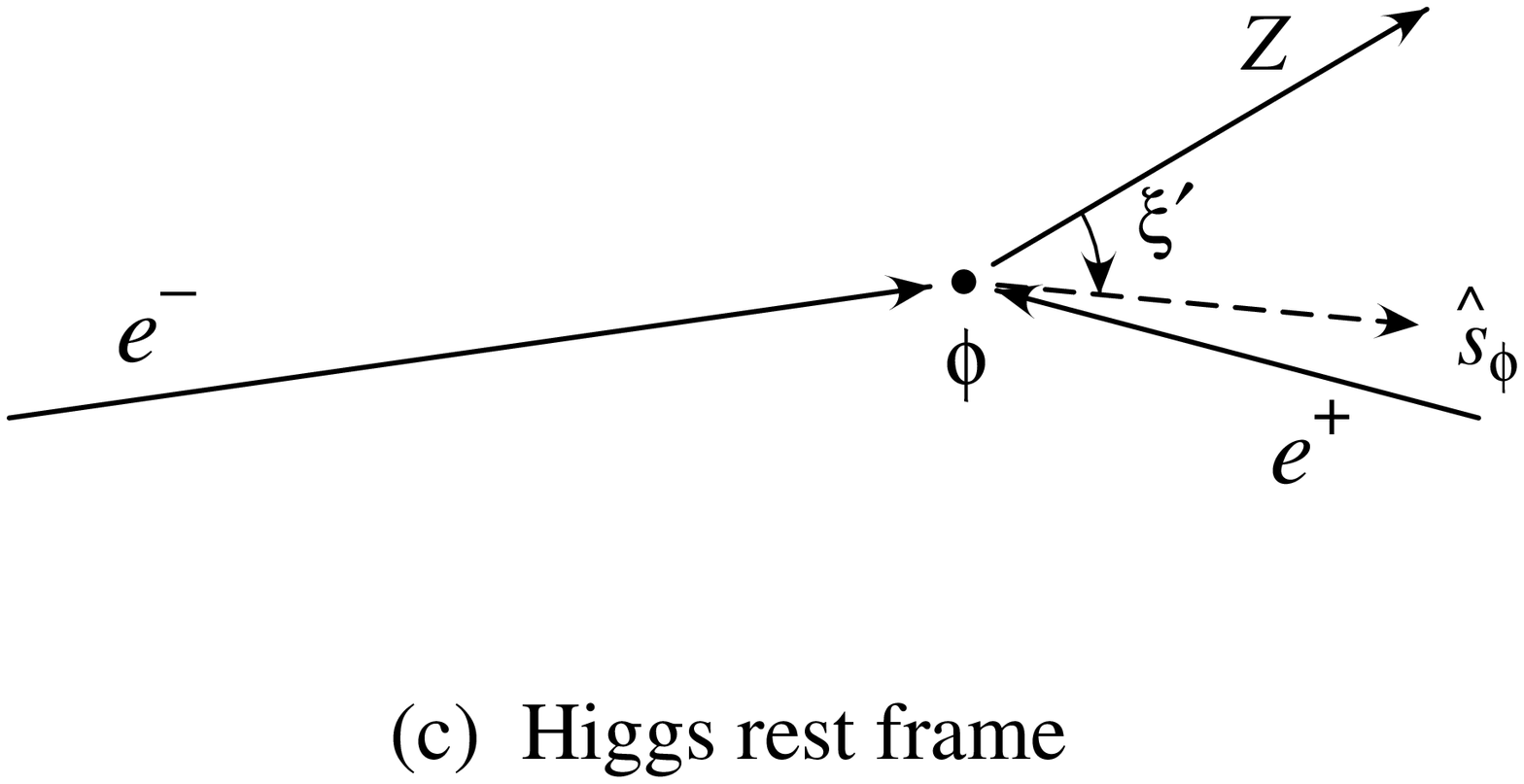}
\vspace{2.0cm}
\caption[]{The scattering process in (a) the zero momentum frame,
(b) the rest-frame of the $Z$ boson and (c) the  rest-frame of the
Higgs.
The spin axis for the $Z$ ($\phi$) is $\hat{s}_{{}_Z}$ ($\hat{s}_{{}_\phi}$).
}
\label{XIdef}
\end{figure*}
To describe the polarized production cross sections for 
$e^{+}e^{-} \longrightarrow Z\phi$ followed by the subsequent
decay of the $Z$ 
we adopt the $\xi$-basis,
introduced by Parke and Shadmi in Ref.~\cite{xibasis}.
As illustrated in Fig.~\ref{XIdef},
the zero momentum frame (ZMF) production angle
$\thetas$ is defined as
the angle between the electron and $Z$ momentum directions.
The spin states for the $Z$ boson
are defined in the its rest frame,
where we decompose its spin along the direction
$\hat{s}_{{}_Z}$, which makes an angle $\xi$ 
in the clockwise direction from the
Higgs momentum.
Although our method does not require detailed observation of the
Higgs boson decay products,
for completeness,
we will go ahead and define a Higgs ``spin'' axis, $\hat{s}_{{}_\phi}$.
This unit vector is located at the angle $\xi'$
in the clockwise direction from the $Z$ boson momentum.
The spin-zero character of the Higgs will be reflected in a lack
of any dependence of the amplitudes on the choice of this axis ($\xi'$).

We denote the two transverse polarization states of the $Z$ boson
by $\a$ and $\b$
(or, equivalently, $(+)$ and $(-)$ respectively)
and the \longitudinal\ state by (0).  
Throughout this paper 
we use the terms ``transverse''
and ``longitudinal''  to refer to directions relative to the
spin axis rather than to the direction of motion of the particle.
A generic vector boson spin will be designated by $\lambda$.
If we sum over all of the polarizations of the 
$Z$ boson, then the dependence on $\xi$ drops out
of the result.

Within this generic 
framework, specific spin bases
are defined by stating the relationship between $\xi$, $\thetas$,
and any other relevant event parameters.  For example,
the familiar helicity basis is defined by fixing
\beq
\xi \equiv \pi.
\label{HELICITYdef}
\eeq
In this case, the spin axes are defined along the directions of
motion of the particles as seen in the ZMF.
Later in this paper, we will encounter additional bases, whose
definitions are inspired by the form of the matrix elements
for the processes under consideration.

Except for the fermion masses, which we set equal to zero,
all input masses and coupling constants used in the computations
presented in this paper are the central values as reported
in the 2004 Review of Particle Properties~\cite{PDB04}.
Consistent with the zero fermion mass approximation, 
we set the coupling between the electron and the Higgs to zero.


\vfill\eject\section{Associated Higgs Production at a Linear Collider}\label{AssocProd}


\subsection{Polarized $Z\HIGGS$ Production}\label{polZHprod}

\begin{figure*}

\vspace*{10.5cm}
\includegraphics{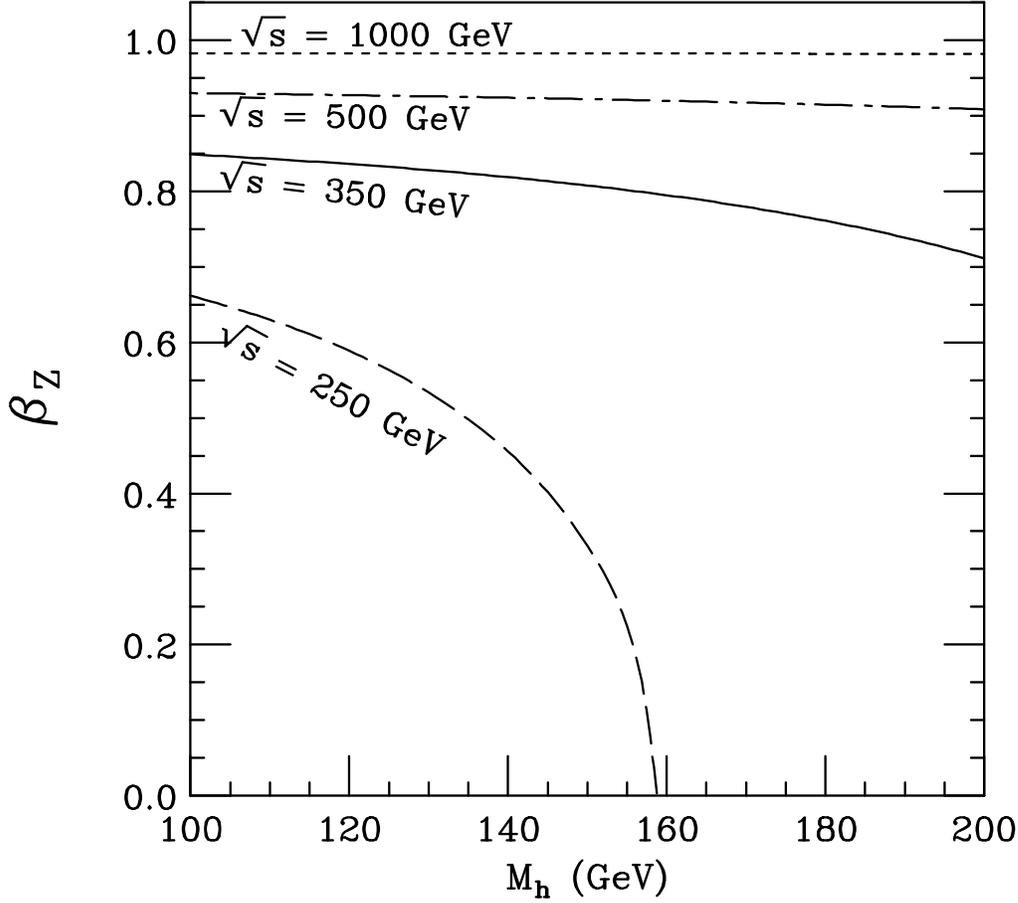}
\vspace{2.0cm}

\caption[]{Zero momentum frame speed of the $Z$ boson, $\bz$,
as a function of the Higgs mass 
in $e^{+} e^{-} \longrightarrow Z\HIGGS$
for collider energies
$\sqrt{s} = 250 \GeV$, $350 \GeV$, $500 \GeV$, and $1000 \GeV$.
}
\label{sbeta}
\end{figure*}

The two particles produced in the process $\eebar\rightarrow Z\HIGGS$,
being of different masses, will have different speeds in the ZMF.
Thus, we may choose to write the amplitudes 
in terms of the ZMF speed
of the $Z$ boson $\bz$,  or
the ZMF speed of the Higgs $\beta_{{}_\HIGGS}$;
we have chosen to use $\bz$. 
In addition to the masses of the Higgs and $Z$ bosons,
the value of $\bz$ also
depends on $\sqrt{s}$, the center-of-mass energy of the collider:
\beq
\BETA =
{ 
\sqrt{[s-(M_{{}_Z}+M_{{}_\HIGGS})^2][s-(M_{{}_Z}-M_{{}_\HIGGS})^2]}
  \over
{s-M_{{}_\HIGGS}^2+M_{{}_Z}^2}
}.
\label{betaZ}
\eeq
A plot of $\BETA$ as a function of the (still-unknown) value of $M_{{}_\HIGGS}$
for various center-of-mass energies
appears in Fig.~\ref{sbeta}.
For reasons of simplicity, we retain both $s$ and $\BETA$
in our expressions rather than using Eq.~(\ref{betaZ}) to eliminate
one of them.

If we neglect the electron mass, there is but a single
diagram for $\eebar\rightarrow Z\HIGGS$, as 
displayed in Fig.~\ref{FeynDiagram}~\cite{assume}.
With the aid of the formalism described in 
Ref.~\cite{WWZZZH},
it is straightforward to calculate the 
differential cross section for $Z\HIGGS$ production using polarized
beams where the
$Z$ is in the spin state $\lambda$:
\beqa
{
{ d\sigma^{\lambda}_L(e_L^{-}e_R^{+}{\rightarrow}Z\HIGGS) }
\over
{ d(\cos\thetas) }
} =  G_F^2  
{{M_W^2 M_Z^2}\over{8\pi s}}
\Theta(s,M_\HIGGS, M_{{}_Z})
\ts q_{eL}^2 \ts  [\T_L^{\lambda}(\BETA,\thetas,\xi)]^2,
\label{ZHsigmaLR}
\eeqa
\beqa
{
{ d\sigma^{\lambda}_R(e_R^{-}e_L^{+}{\rightarrow}Z\HIGGS) }
\over
{ d(\cos\thetas) }
} =  G_F^2  
{{M_W^2 M_Z^2}\over{8\pi s}}
\Theta(s,M_\HIGGS, M_{{}_Z})
\ts q_{eR}^2 \ts [\T_R^{\lambda}(\BETA,\thetas,\xi)]^2.
\label{ZHsigmaRL}
\eeqa
In these expressions,
$G_F$ is the Fermi coupling constant, $M_W$ is the mass
of the $W$ boson, and $\theta_W$ is the
Weinberg angle.
The kinematics associated with the threshold behavior of
the cross section have been collected into the function
$\Theta(s,M_\HIGGS,M_{{}_Z})$, which is defined to be
\beq
\Theta(s,M_\HIGGS, M_Z) \equiv
2\BETA\GAMMA\ts
{ {M_{{}_Z}} \over {\sqrt{s}}   } \ts
\Biggl(
{  {s+M_{{}_Z}^2-M_{{}_\HIGGS}^2} \over {s-M_{{}_Z}^2}  }
\Biggr)^2.
\label{ThreshFun}
\eeq
Here $\sqrt{s}$ is the collider center-of-mass energy,
$\BETA$ the speed of the $Z$ boson in the zero momentum frame (ZMF)
of the event, and $\GAMMA$ the usual relativistic boost factor,
\beq
\GAMMA \equiv (1-\BETA^2)^{-1/2}.
\eeq  
We have chosen the factors that comprise 
$\Theta(s,M_\HIGGS, M_Z)$ so that for $\BETA\longrightarrow 1$
(equivalently, $s\longrightarrow\infty$) we have
$\Theta\longrightarrow 1$.  Naturally, $\Theta = 0$ at threshold.
We parameterize the $Zf\bar{f}$ vertex as
\beq
i\Gamma^\mu \equiv
ig\Bigl[ q_{fL} \gamma^{\mu}\half(1-\gamma_5)
+
 q_{fR} \gamma^{\mu}\half(1+\gamma_5)\Bigr].
\eeq
Here $g$ is the weak coupling constant; it is connected to the Fermi
coupling constant in the usual manner:
\beq
{ {G_F}\over{\sqrt{2}} } =
{ {g^2}\over{8M_{{}_W}^2} }.
\eeq
The quantity $q_{fL}$ describes the coupling of a left-handed 
fermion line of flavor $f$ to the $Z$ boson while 
$q_{fR}$ represents the coupling of the $Z$ boson
to a right-handed fermion line.  These couplings
are collected in Table~\ref{CouplingTable} for easy reference.
\begin{table*}
\caption{Standard Model couplings of fermions to the $Z$ boson.
The numerical values correspond to $\SSQW=0.2312$.
\label{CouplingTable}}
\begin{ruledtabular}
\begin{tabular}{cccccccc}
&& fermion &   $q_{fL}$  &&   $q_{fR}$  && \\[0.05in]
\colrule
&& $u,c$ &    
   $ \displaystyle{ {(1-{4\over3}\SSQW)}\over{2\cos\theta_W} } = .3945 $
&$\phantom{\Biggl[}$&
   $ \displaystyle{ {({-{4\over3}}\SSQW)}\over{2\cos\theta_W} }=-.1758 $
\\[0.13in]        
&& $d,s,b$ &  
   $ \displaystyle{ {(-1+{2\over3}\SSQW)}\over{2\cos\theta_W} }=-.4824 $
&&
   $ \displaystyle{ {({{2\over3}}\SSQW)}\over{2\cos\theta_W} }=.0879 $ 
\\[0.13in]        
&& $e,\mu,\tau$ & 
   $ \displaystyle{ {(2\SSQW-1)}\over{2\cos\theta_W} }=-.3066 $
&&
   $ \displaystyle{ {(2\SSQW)}\over{2\cos\theta_W} }=.2637 $ 
\\[0.13in]        
&& $\nu$ &       
   $ \displaystyle{ {1}\over{2\cos\theta_W} }=0.5702 $
   && $0$ && \\[0.10in]
\end{tabular}
\end{ruledtabular}
\end{table*}
All of the remaining spin information
is contained in the spin functions, $\T_{L,R}^\lambda$, 
which, 
for $e^{-}_L e^{+}_R$,  
are given by ($\sth \equiv \sin\theta^*$, 
$\cth \equiv \cos\theta^*$, etc.):
\beqa
\T_L^{\pm}(\BETA,\thetas,\xi) 
&=&  {1\over{\sqrt{2}}} \Bigl[ \sth\sxi + \ginv(\cth\cxi \pm 1) \Bigr];
\cr
\T_L^{0}(\BETA,\thetas,\xi) 
&=& \ginv\cth\sxi - \sth\cxi.
\phantom{\biggl[}
\label{ZH-LR}
\eeqa
The spin functions for a right-handed
electron line are related to the left-handed functions via
\beqa
\T_R^{\pm}(\BETA,\thetas,\xi) &=& \T_L^{\mp}(\BETA,\thetas,\xi)
\cr
\T_R^{0}(\BETA,\thetas,\xi) &=& \T_L^{0}(\BETA,\thetas,\xi).
\label{ZH-RL}
\eeqa

To provide some sense of how the polarized production amplitudes
depend on our choice of spin basis we now examine the explicit
form of these amplitudes in a couple of cases of interest.
First, in the helicity basis ($\xi = \pi$) we have
\beqa
\T_L^{\pm}(\BETA,\thetas,\pi) 
&=&  -{\ginv\over{\sqrt{2}}} (\cth \mp 1)  
\cr
\T_L^{0}(\BETA,\thetas,\pi) 
&=&  \sth.
\phantom{\biggl[}
\label{ZH-helicity}
\eeqa
Unless $\BETA$ is rather close to 1, one consequence of these expressions
is that a non-negligible fraction of the total integrated
cross section 
will be supplied by each of the three 
spins (see Fig.~\ref{ZHbetaplot} below).
A second consequence of the form of these amplitudes is the
equality of the contributions to the total integrated cross
section from the
$(+)$ and $(-)$ spin components at all values $\BETA$,
even for polarized beams (odd functions of $\cth$ drop out
when we integrate over $\cos\theta^{*}$; 
the only difference
between $\vert\T_L^+\vert^2$ and $\vert\T_L^-\vert^2$ 
is in the sign of the $\cos\theta^{*}$ cross term).

We contrast these results with the $Z\HIGGS$-transverse basis,
which was introduced in Ref.~\cite{WWZZZH}.
This basis was motivated by a 
desire to eliminate one of the three spin components.
Now, according to Eqs.~(\ref{ZH-LR})
and~(\ref{ZH-RL}), it is not possible to make both of the
spin functions $\T_L^\lambda$ and $\T_R^\lambda$
vanish simultaneously when $\lambda$ is $(+)$ or $(-)$.  
Consequently, 
for unpolarized beams, it is impossible to choose a spin
basis for which either variety of transversely-polarized $Z$ boson
is absent.
On the other hand, it is possible to eliminate
the contribution from the
longitudinal $Z$~bosons by choosing
\beq
\sxi = {{\sth}\over{\sqrt{1-\bz^2\cth^2}}};
\qquad
\cxi = {{\ginv\cth}\over{\sqrt{1-\bz^2\cth^2}}},
\label{magic-xi}
\eeq
which may be abbreviated as
\beq
\tan\xi = \GAMMA \tan\theta^{*}.
\label{magic-xi-tangent}
\eeq
Eq.~(\ref{magic-xi}) defines the  $Z\HIGGS$-transverse
basis, so-called because in this basis only $Z$'s with transverse
polarizations are produced.  Although the choice of $\xi$ implied
by Eq.~(\ref{magic-xi}) looks like a radical departure from the
simplicity ($\sxi = 0$; $\cxi = -1$) of the ZMF helicity basis 
in that it selects a different spin axis for each event, in reality, 
the same is true for the choice $\xi = \pi$.  Recall that
$\xi$ is defined relative to the direction of motion of the $Z$ boson
in the ZMF: this direction is different for each event.
What we have done in defining the $Z\HIGGS$-transverse basis is to include
some cleverly-selected $\bz$-dependence in addition to the 
(now explicit) $\thetas$-dependence.  Note that the existence of
this basis depends on neither the machine energy nor
the Higgs mass:  Eqs.~(\ref{betaZ}) and~(\ref{magic-xi}) remain
well-defined so long as $\sqrt{s} \ge M_{{}_Z} + M_{{}_\HIGGS}$
({\it i.e.}\ a $Z\HIGGS$ final state must be kinematically allowed).  
In particular, for $\BETA \rightarrow 1$, we have
\beq
\sxi \rightarrow 1; \qquad \cxi \rightarrow 0,
\label{ZHhighE}
\eeq
that is, $\xi \rightarrow \pi/2$.  This is clearly {\it not}\ the
helicity basis,  
suggesting that even far above threshold
the \ZHtrans\ basis represents a different and potentially interesting
way of viewing the data. 
At the other energy extreme, $\BETA \rightarrow 0$, we have
\beq
\sxi \rightarrow \sth; \qquad \cxi \rightarrow \cth.
\label{ZHthresh}
\eeq
A moment's consideration of Fig.~\ref{XIdef} along with 
Eq.~(\ref{ZHthresh}) will lead to the recognition that the
directions of the incoming beams are being used to decompose the
spins in this limit.  That is, at threshold, the $Z\HIGGS$-transverse
basis is coincident with the so-called beamline basis\cite{ttbar1}
(see Appendix~\ref{BML}).

In addition to eliminating the contribution from the longitudinal
spin component, 
the \ZHtrans\ basis is also the basis in which the $+$ and $-$
components are each maximized. 

In the \ZHtrans\ basis, the explicit forms of the spin functions
are
\beqa
\T_L^{\pm}\Bigl(\BETA,\theta^*,\tan^{-1}(\GAMMA\tan\theta^*)\Bigr) & = & 
{{{1}}\over{\sqrt{2}}}  
\biggl( \sqrt{1-\BETA^2\cos^2\theta^*}\pm\sqrt{1-\BETA^2} \ts\biggr)
\cr
\T_L^0\Bigl(\BETA,\theta^*,\tan^{-1}(\GAMMA\tan\theta^*)\Bigr) & = & 0.
\label{ZH-ZHtrans}
\eeqa
It is clear from these expressions that the $(+)$ and $(-)$ states
are equally-populated for polarized beams only in the ultra-relativistic
limit ($\BETA\rightarrow1$); for small $\BETA$, one of these
two states is approximately empty.

\begin{figure*}

\vspace*{13cm}
\includegraphics{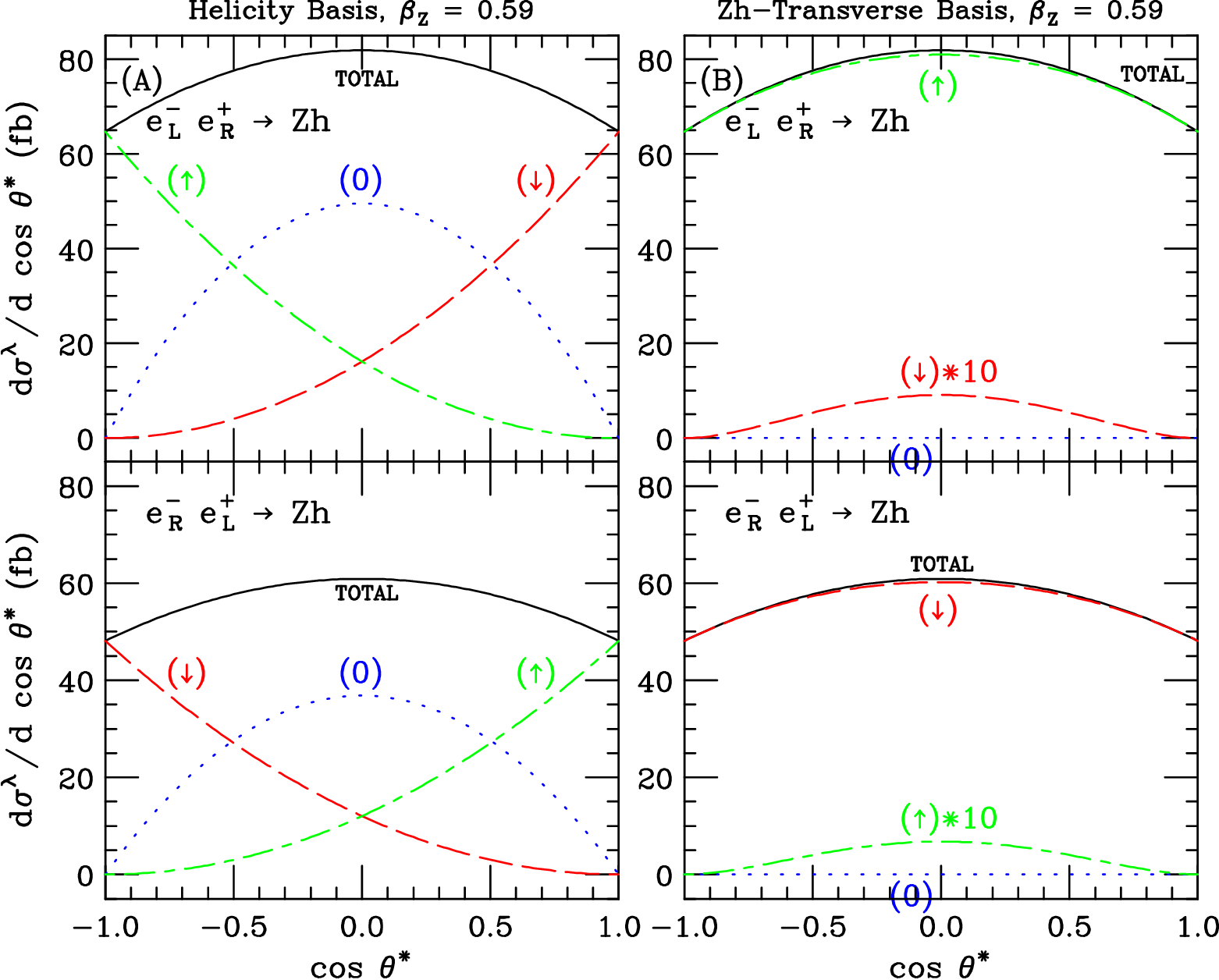}

\caption[]{Production angle distributions for the polarized
$\eebar\rightarrow Z\HIGGS$ cross section 
assuming $\sqrt{s} = 250$ GeV and $M_h = 120$ GeV 
($\beta_{{}_{Z}}=0.59$).   
Displayed are
the contributions from the three possible $Z$ spins in 
{\bf (A)} the helicity basis and 
{\bf (B)}\ the \ZHtrans\ basis. 
}
\label{ZH-CTH}
\end{figure*}

In Fig.~\ref{ZH-CTH} we display plots of the contributions
from the three spin states as a function of the $Z$ production
angle $\cos\theta^*$ at $\BETA=0.59$, corresponding to
the not-implausible combination $m_{{}_\HIGGS} = 120\GeV$,
$\sqrt{s}=250\GeV$.  These plots clearly exhibit the features noted
above:  in the helicity basis, all three spin components make 
significant contributions to the total cross section whereas
in the \ZHtrans\ basis a single component dominates depending on
the polarization of the incoming beams.

Given a choice of basis (i.e. $\xi$) and machine energy (i.e. $\BETA$,
once $M{{}_\HIGGS}$ is known)
we may integrate Eqs.~(\ref{ZHsigmaLR}) and~(\ref{ZHsigmaRL}) 
over $\cos\theta^*$
to determine the fraction of $Z$ bosons produced in each of the
three possible spin states.
In Fig.~\ref{ZHbetaplot} we present the results as a function
of $\BETA$;
Table~\ref{ZH-Breakdowns} lists the numerical values
corresponding to $\bz = 0.59$, that is, at the same Higgs mass 
and machine energy
considered in Fig.~\ref{ZH-CTH}. The fractions in the helicity basis 
are essentially
as anticipated above, starting at an equal mixture of $(+)$, $(-)$
and $(0)$ at threshold and becoming 100\% longitudinal 
as $\BETA\rightarrow 1$.  At all energies the fractions of $(+)$ 
and $(-)$ spins are equal, even for polarized beams.  In
contrast, the \ZHtrans\ basis fractions are relatively insensitive
to $\BETA$ (unless $\BETA$ gets fairly large).

\begin{table*}
\caption{Spin decompositions in selected bases for 
$\eebar \rightarrow Z\HIGGS$
assuming $m_h = 120 \GeV$ and $\sqrt{s} = 250 \GeV$
($\beta_{{}_Z}=0.59$).
\label{ZH-Breakdowns}}
\begin{ruledtabular}
\begin{tabular}{crrr}
 & $(+)$ &  $(-)$ &  $(0)$  \\[0.05in]
\hline
helicity basis: &&& \\
\hline
$e_L^- e_R^+$            & 28.3\% & 28.3\% &  43.4\%  \\
$e_R^- e_L^+$            & 28.3\% & 28.3\% &  43.4\%  \\
$e^- e^+$ (unpolarized)  & 28.3\% & 28.3\% &  43.4\%  \\
\hline
$Z\HIGGS$-transverse basis: &&& \\
\hline
$e_L^- e_R^+$           & 99.3\% & 0.7\% &  0.0\%\footnotemark[1]  \\
$e_R^- e_L^+$           & 0.7\% & 99.3\% &  0.0\%\footnotemark[1]  \\
$e^- e^+$ (unpolarized) & 57.3\% & 42.7\% &  0.0\%\footnotemark[1]  \\
\end{tabular}
\end{ruledtabular}
\footnotetext[1] {This contribution is exactly zero (by construction).}
\end{table*}

\begin{figure*}

\vspace*{13cm}
\includegraphics{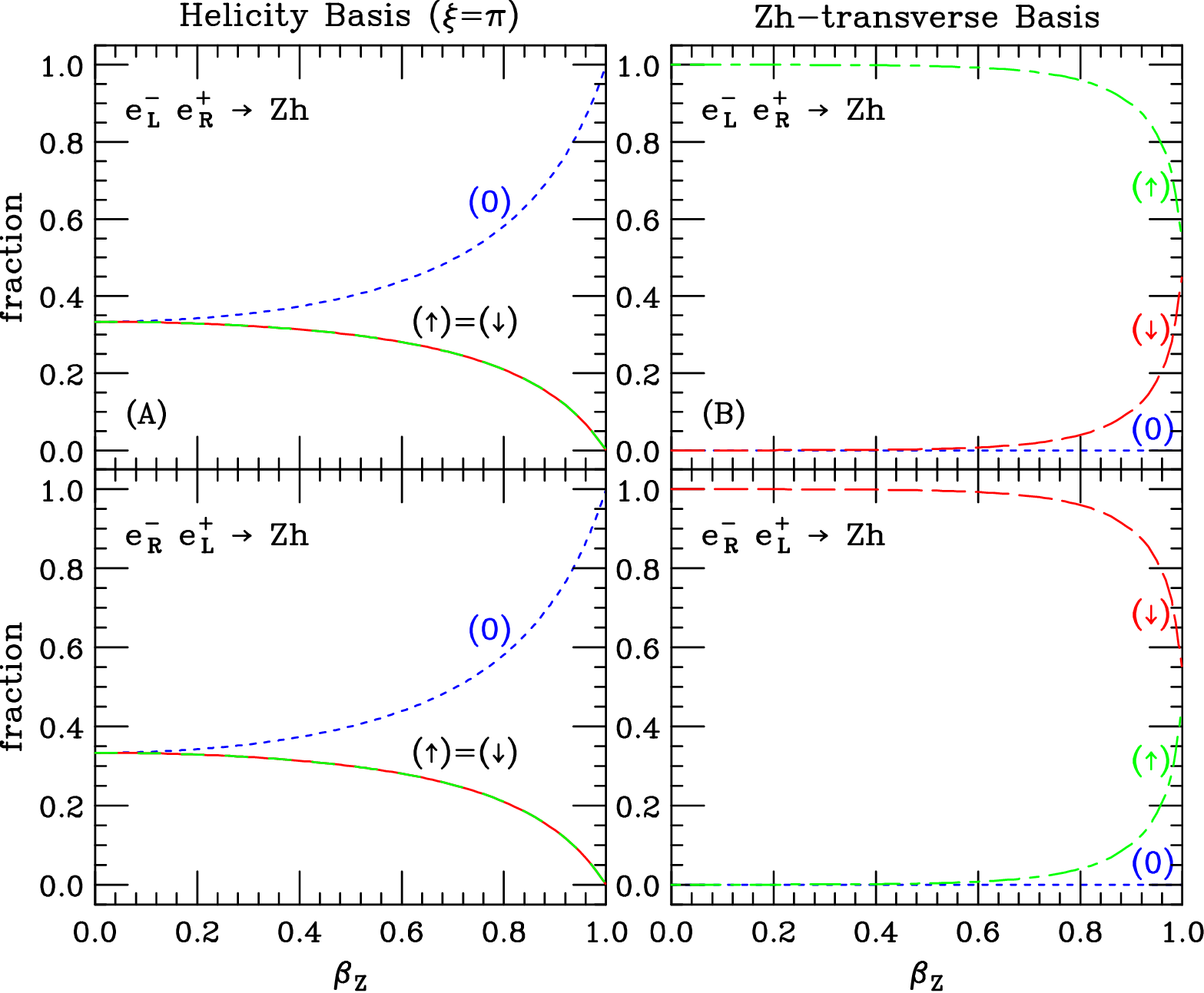}

\caption[]{Spin decomposition in the {\bf (A)}
helicity and 
{\bf (B)}
$Z\HIGGS$-transverse bases
of the $e_L^{-}e_R^{+}\rightarrow Z\HIGGS$
and $e_R^{-}e_L^{+}\rightarrow Z\HIGGS$
cross sections as a function of the ZMF speed $\bz$ of the $Z$
boson.
Shown are the fractions of
the total cross section in the $(\uparrow)$, $(\downarrow)$, 
and $(0)$ spin states.
}
\label{ZHbetaplot}
\end{figure*}

Summing over the three possible spins of the $Z$ 
we obtain the total differential cross section
\beqa
\sum_{\lambda}
{
{ d\sigma_L^{\lambda} }
\over
{ d(\cos\thetas) }
} =  q_{eL}^2 G_F^2 
{{M_W^2 M_Z^2}\over{8 \pi s}}
\Theta(s, M_\HIGGS, M_{{}_Z})
\biggl[
2-\BETA^2(1+\cos^2\thetas)
\biggr].
\label{ZHsigmaDIFF-LR}
\eeqa
If we replace $q_{eL}$ by $q_{eR}$ in Eq.~(\ref{ZHsigmaDIFF-LR})
we obtain the result for $e_R^{-} e_L^{+}$ scattering.
Eq.~(\ref{ZHsigmaDIFF-LR}) 
is independent of the spin axis angle $\xi$, as it must be.
This expression corresponds to the curves labeled ``TOTAL''
in Fig.~\ref{ZH-CTH}.

By integrating Eq.~(\ref{ZHsigmaDIFF-LR}) 
over $\cos\thetas$
we obtain the total cross section for 
$e_L^{-}e_R^{+}\longrightarrow Z\HIGGS$:
\beq
\sigma_L( e_L^{-}e_R^{+}\longrightarrow Z\HIGGS) = 
 q_{eL}^2 G_F^2  
{{M_W^2 M_Z^2}\over{2 \pi s}}
\Theta(s,M_\HIGGS, M_{{}_Z})
\Bigl[
1- \hbox{$2\over3$}\BETA^2
\Bigr].
\label{ZHsigmaTOTAL-LR}
\eeq
Once again, the result for $e_R^- e_L^+$ scattering
may be generated by the replacement $q_{eL}\rightarrow q_{eR}$.
Finally, we may divide the differential 
cross section~(\ref{ZHsigmaDIFF-LR})
by the corresponding total cross section to obtain 
the $Z\HIGGS$ production cross section for
polarized beams normalized to unity:
\beqa
{{1}\over{\sigma_L}}\ts
{{d\sigma_L}\over{d(\cos\thetas)}}
&=& {{3}\over{4}}\ts{{2-\bz^2(1+\cos^2\thetas)}
                   \over
                   {3-2\bz^2}}
\cr &=&
{{1}\over{\sigma_R}}\ts
{{d\sigma_R}\over{d(\cos\thetas)}}.
\label{ZHunityLR}
\eeqa
Eq.~(\ref{ZHunityLR}) explicitly exhibits the fact that
the only difference between $e_L^{-}e_R^{+}$
and $e_R^{-}e_L^{+}$ scattering to the $Z\HIGGS$
final state is in the overall normalization of the 
production cross section:  the shapes of the two distributions
are identical.  Put another way, the differential cross-section
for $e_{R}^{-}e_{L}^{+}\rightarrow Z\HIGGS$ may be obtained 
by rescaling the cross section for
$e_{L}^{-}e_{R}^{+}\rightarrow Z\HIGGS$ by the
factor $q_{eR}^2/q_{eL}^2 = 0.7397$.


\subsection{Polarized $ZA$ Production}\label{pseudo}

In the Standard Model, the minimal Higgs sector consists of a
single complex doublet before spontaneous-symmetry breaking,
leading to
a physical spectrum that contains only a
$CP$-even Higgs boson.
In extensions to the Standard Model, the Higgs sector is often
more complicated, leading to additional physical Higgs states
with various quantum numbers.  
One alternative which occurs frequently is the
so-called pseudoscalar Higgs boson $A^0$ (strictly-speaking,
the $A^0$ is a $CP$-odd Higgs;
its ${\cal J}^{PC}$ assignment is $0^{+-}$).
For example, the minimal supersymmetric standard model (MSSM)
contains such a state in addition to a pair of $0^{++}$ states
$h^0$ and $H^0$.
Thus, it is interesting to consider
how the 
angular distributions are different for a pseudoscalar 
Higgs boson.

In order to couple a $CP$-odd Higgs boson to a pair of vector
bosons in a gauge-invariant manner
that conserves $CP$, it is necessary for the interaction to
take the form~\cite{HH}
\beq
\epsilon_{\mu\nu\rho\sigma} F^{\mu\nu} F^{\rho\sigma} A^0.
\label{ZZAcoupling}
\eeq
This is a dimension-5 operator; it cannot appear in the tree-level
Lagrangian.  Such a coupling can be generated at loop level, however.
Thus, as in
Ref.~\cite{Barger}, we write the (effective) $ZZA$ vertex as
\beq
-{ {igM_{{}_Z}}\over{\cos\theta_W} }
{ {\eta}\over{\Lambda^2} }
k_1^{\mu} k_2^{\nu} \varepsilon_{\mu\nu\alpha\beta},
\label{ZZAvertex}
\eeq
where $k_1$ and $k_2$ are the 4-momenta of the two $Z$'s,
$\eta$ is a dimensionless coupling constant, and $\Lambda$
is the mass scale at which this vertex is generated~\cite{Lambda}.
Because of the generality of the argument leading to the form 
of the coupling in Eqs.~(\ref{ZZAcoupling}) and~(\ref{ZZAvertex}),
the angular correlations in a wide variety of models will have
this form even though we cannot say very much about the over-all
size of the total cross section.
Nevertheless, we expect 
$e^{+} e^{-} \longrightarrow Z \HIGGS$ (if not kinematically suppressed)
to dominate over $e^{+} e^{-} \longrightarrow Z A$ on the grounds
that
the one-loop effective $ZZA$ coupling will likely be smaller than
the tree-level $ZZ\HIGGS$ coupling.

The vertex in Eq.~(\ref{ZZAvertex}) leads to the differential 
cross sections
\beq
{
{ d\sigma^{\lambda}(e_L^{-}e_R^{+}{\rightarrow}ZA) }
\over
{ d(\cos\thetas) }
} =  G_F^2 M_W^2 \eta^2
 {{M_{{}_Z}^4} \over{\Lambda^4} }  \ts
{ {\BETA^2}\over{\pi } }
\Theta(s,M_{{}_A}, M_{{}_Z}) \ts
q_{eL}^2 \ts
\Bigl[
 \widetilde\T_L^{\lambda}(\thetas,\xi)\Bigr]^2
\label{ZAsigma-LR}
\eeq
and
\beq
{
{ d\sigma^{\lambda}(e_R^{-}e_L^{+}{\rightarrow}ZA) }
\over
{ d(\cos\thetas) }
} = G_F^2 M_W^2 \eta^2
{M_{{}_Z}^4\over\Lambda^4}  \ts
{ {\BETA^2}\over{\pi} }
\Theta(s,M_{{}_A}, M_{{}_Z}) \ts
q_{eR}^2 \ts
\Bigl[
 \widetilde\T_R^{\lambda}(\thetas,\xi)\Bigr]^2
\label{ZAsigma-RL}
\eeq
In this case, the spin functions turn out to be independent of energy:
\beqa
\widetilde\T_L^{\pm}(\thetas,\xi) 
&=&  {1\over\sqrt{2}} \bigl( \cth\pm \cxi \bigr) ; \cr
\widetilde\T_L^{0}(\thetas,\xi)
&=&  \sxi.
\label{ZA-LR}
\eeqa
The spin functions for $e_R^{-} e_L^{+}$ scattering are
related to those for $e_L^{-} e_R^{+}$ in the usual fashion, 
namely
\beqa
\widetilde\T_R^{\pm}(\thetas,\xi) &=& 
\widetilde\T_L^{\mp}(\thetas,\xi); \cr
\widetilde\T_R^{0}(\thetas,\xi) &=& 
\widetilde\T_L^{0}(\thetas,\xi).
\label{ZA-RL}
\eeqa

It is obvious from
the especially simple form of the spin functions in
Eqs.~(\ref{ZA-LR}) and~(\ref{ZA-RL})
that the optimal basis for studying angular correlations
in $ZA$ production and decay is the helicity basis, 
independent of the machine energy.   In the helicity basis,
the spin functions become
\beqa
\widetilde\T_L^{\pm}(\thetas,\pi) 
&=&  {1\over\sqrt{2}} \bigl( \cth\mp 1 \bigr) ; \cr
\widetilde\T_L^{0}(\thetas,\pi)
&=&  0.
\label{ZA-helicity}
\eeqa
Only the helicity basis has 
the property
that one of the three amplitudes (the longitudinal one) vanishes.
\eject

\begin{figure*}

\vspace*{13cm}
\includegraphics{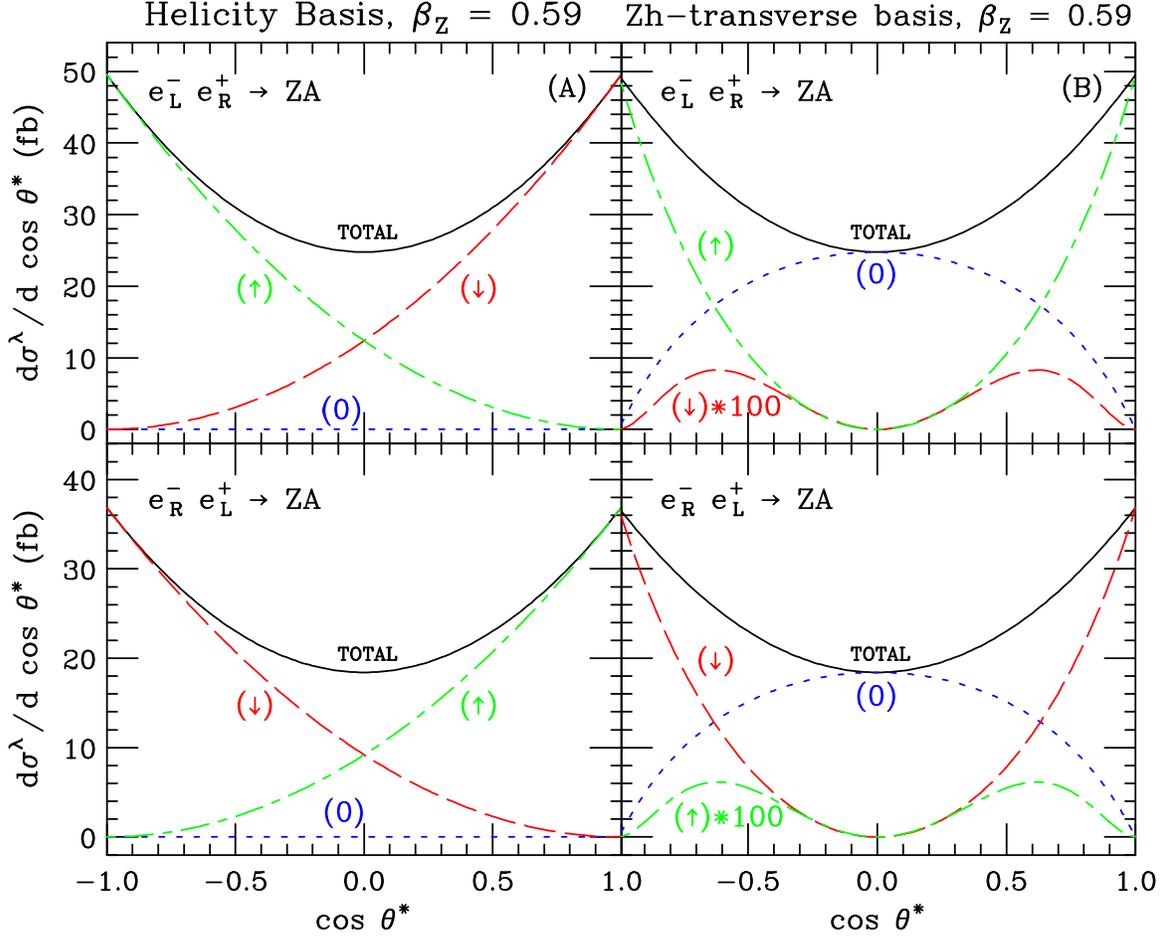}

\caption[]{Production angle distributions for the polarized
$\eebar\rightarrow ZA$ cross section 
assuming $\sqrt{s} = 250$ GeV and $M_h = 120$ GeV 
($\beta_{{}_{Z}}=0.59$)   
broken down
into the contributions from the three possible $Z$ spins, in the
{\bf (A)}\
helicity and 
{\bf (B)}\
\ZHtrans\ bases. 
}
\label{ZA-CTH-hi}
\end{figure*}

For the sake of comparison, it is useful to write out the
$ZA$ spin functions in the \ZHtrans\ basis:
\beqa
\widetilde\T_L^{\pm}\biggl(\thetas,
\tan^{-1}(\GAMMA\tan\theta^*)\biggr) 
&=&  {{\cth}\over\sqrt{2}} 
{
{\sqrt{1-\BETA^2\cth^2}\pm \sqrt{1-\BETA^2} }
\over
{\sqrt{1-\BETA^2\cth^2}}
} ; \cr
\widetilde\T_L^{0}\biggl(\thetas,
\tan^{-1}(\GAMMA\tan\theta^*)\biggr)
&=&  {{\sth}\over{\sqrt{1-\BETA^2\cth^2}}}.
\label{ZA-ZHtransverse}
\eeqa
The results are
not particularly simple since this choice of $\xi$ was concocted
to simplify the $Z\HIGGS$ amplitudes, not the $ZA$ amplitudes.
We display the $\cos\theta^*$-dependence of these contributions to
the $ZA$ production cross section in Fig.~\ref{ZA-CTH-hi}
for the same Higgs mass and machine energy as previously.
Perhaps surprisingly, one of the spin states is
highly suppressed in the 
\ZHtrans\ basis when
we use polarized beams.  This
is in contrast to the helicity basis, where  
the ($\uparrow$) and
($\downarrow$) states provide equal contributions to the
total cross section.

\begin{figure*}

\vspace*{13cm}
\includegraphics{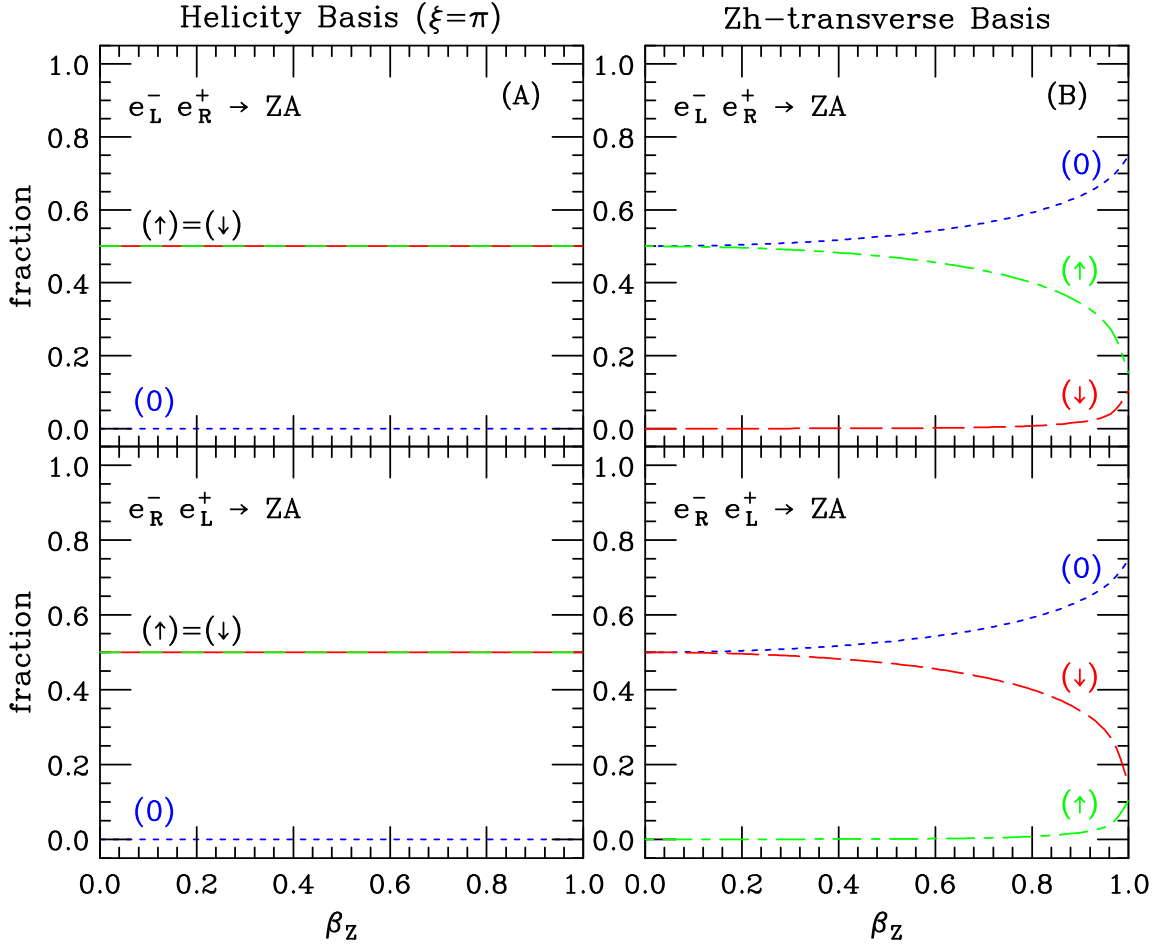}

\caption[]{Spin decomposition in the {\bf (A)}
helicity and {\bf (B)}
$Z\HIGGS$-transverse bases
of the $e_L^{-}e_R^{+}\rightarrow ZA$
and $e_R^{-}e_L^{+}\rightarrow ZA$
cross sections as a function of the ZMF speed $\bz$ of the $Z$
boson.
Shown are the fractions of
the total cross section in the $(\uparrow)$, $(\downarrow)$, 
and $(0)$ spin states.
}
\label{ZAbetaplots}
\end{figure*}

In Fig.~\ref{ZAbetaplots} we plot the $Z$ spin decomposition
in $e^{+}e^{-}\rightarrow ZA$ production as a function of
$\bz$.  Table~\ref{ZA-Breakdowns} lists selected 
numerical values for this spin breakdown.
\begin{table*}
\caption{Spin decompositions in selected bases for 
$\eebar \rightarrow ZA$
assuming $m_h = 120 \GeV$ and $\sqrt{s} = 250 \GeV$
($\beta_{{}_Z}=0.59$).
\label{ZA-Breakdowns}}
\begin{ruledtabular}
\begin{tabular}{crrr}
 & $(+)$ &  $(-)$ &  $(0)$  \\[0.05in]
\hline
helicity basis: &&& \\
\hline
$e_L^- e_R^+$           & 50.0\% & 50.0\% &  0.0\%  \\
$e_R^- e_L^+$           & 50.0\% & 50.0\% &  0.0\%  \\
$e^- e^+$ (unpolarized) & 50.0\% & 50.0\% &  0.0\% \\
\hline
$Z\HIGGS$-transverse basis: &&& \\
\hline
$e_L^- e_R^+$            & 45.7\% & 0.1\% &  54.1\% \\
$e_R^- e_L^+$            & 0.1\% & 45.7\% &  54.1\% \\
$e^- e^+$ (unpolarized)  & 26.3\% & 19.6\% &  54.1\% \\
\end{tabular}
\end{ruledtabular}
\end{table*}
A consequence of the energy-independence of the spin functions
contained in Eqs.~(\ref{ZA-LR}) and~(\ref{ZA-RL}) is that 
all of the $\bz$-dependence exhibited in these plots comes
from choosing $\xi$ to be an explicit function of $\bz$.
In the helicity basis, the spin breakdown is 50-50 between
the $\a$ and $\b$ spin states for all center-of-mass energies.
On the other hand, in the (non-optimal for $ZA$ production)
\ZHtrans\ basis, we have a 50-50 mix of the $(0)$ and $\a$
or $(0)$ and $\b$ spin states near threshold, with the fraction
of longitudinal $Z$'s gradually increasing as $\BETA\rightarrow 1$.
Thus, we are left with the linguistically-awkward situation where
in the \ZHtrans\ basis, $ZA$ production is dominated by the
{\it longitudinal}\ component! 

Summing over the possible spins of the $Z$ we find that
the total differential cross section corresponding
to the spin functions in Eq.~(\ref{ZA-LR}) reads
\beq
\sum_{\lambda}
{
{ d\sigma_L^{\lambda} }
\over
{ d(\cos\thetas) }
} =  q_{eL}^2 G_F^2 M_W^2 \eta^2
 {M_{{}_Z}^4\over\Lambda^4}  \ts
{{\BETA^2}\over{\pi }}
\Theta(s,M_{{}_A}, M_{{}_Z}) \ts
\Bigl(
1 + \cos^2\thetas
\Bigr).
\label{ZAsigmaDIFF-LR}
\eeq
The corresponding result for $e_R^- e_L^+$ scattering
differs only by the replacement $q_{eL}^2 \rightarrow q_{eR}^2$.
Integrating over $\cos\thetas$
gives the total polarized $ZA$ production cross section
\beq
\sigma_L(e^{+}e^{-}\longrightarrow ZA) 
= {{8}\over{3}}\ts q_{eL}^2 G_F^2 M_W^2 \eta^2
 {M_{{}_Z}^4 \over\Lambda^4 }  \ts
{ {\BETA^2}\over{\pi } }
\Theta(s,M_{{}_A}, M_{{}_Z}).
\label{ZAsigmaTOTAL-LR}
\eeq
The result for $\sigma_R(e^{+}e^{-}\longrightarrow ZA)$ follows
from Eq.~(\ref{ZAsigmaTOTAL-LR}) by the replacement
$q_{eL}^2\rightarrow q_{eR}^2$.
Thus, the normalized differential distributions are
\beqa
{{1}\over{\sigma_L}}\ts
{{d\sigma_L}\over{d(\cos\thetas)}}
&=& {{3}\over{8}}\ts(1 + \cos^2\thetas)
\cr &=& 
{{1}\over{\sigma_R}}\ts
{{d\sigma_R}\over{d(\cos\thetas)}}.
\label{ZAsigmaUNITY-LR}
\eeqa


\vfill\eject\section{Polarized $Z$ Decays}\label{polZdec}

The $Zf\bar{f}$ coupling violates both parity and flavor
universality.  Thus, the angular distributions for the decay
of polarized $Z$ bosons are forward-backward asymmetric, and
depend on which fermions appear in the final state.
Neglecting the mass of the final state fermions~\cite{messy},
the angular distributions in the rest
frame of the decaying $Z$ may be written as
\beq
{ {1}\over{\Gamma_f} }
{ {d\Gamma^{\lambda}} \over {d(\cos\chi)} }
= {3\over 8}
\Biggl\{ 
{{q_{fL}^2}\over{q_{fL}^2+q_{fR}^2}}
\vert \DEE_{L}^{\lambda} \vert^2
+ 
{{q_{fR}^2}\over{q_{fL}^2+q_{fR}^2}}
\vert \DEE_{R}^{\lambda} \vert^2
\Biggr\}.
\label{Zdecays}
\eeq

\begin{figure*}
\vspace*{10.75cm}
\includegraphics{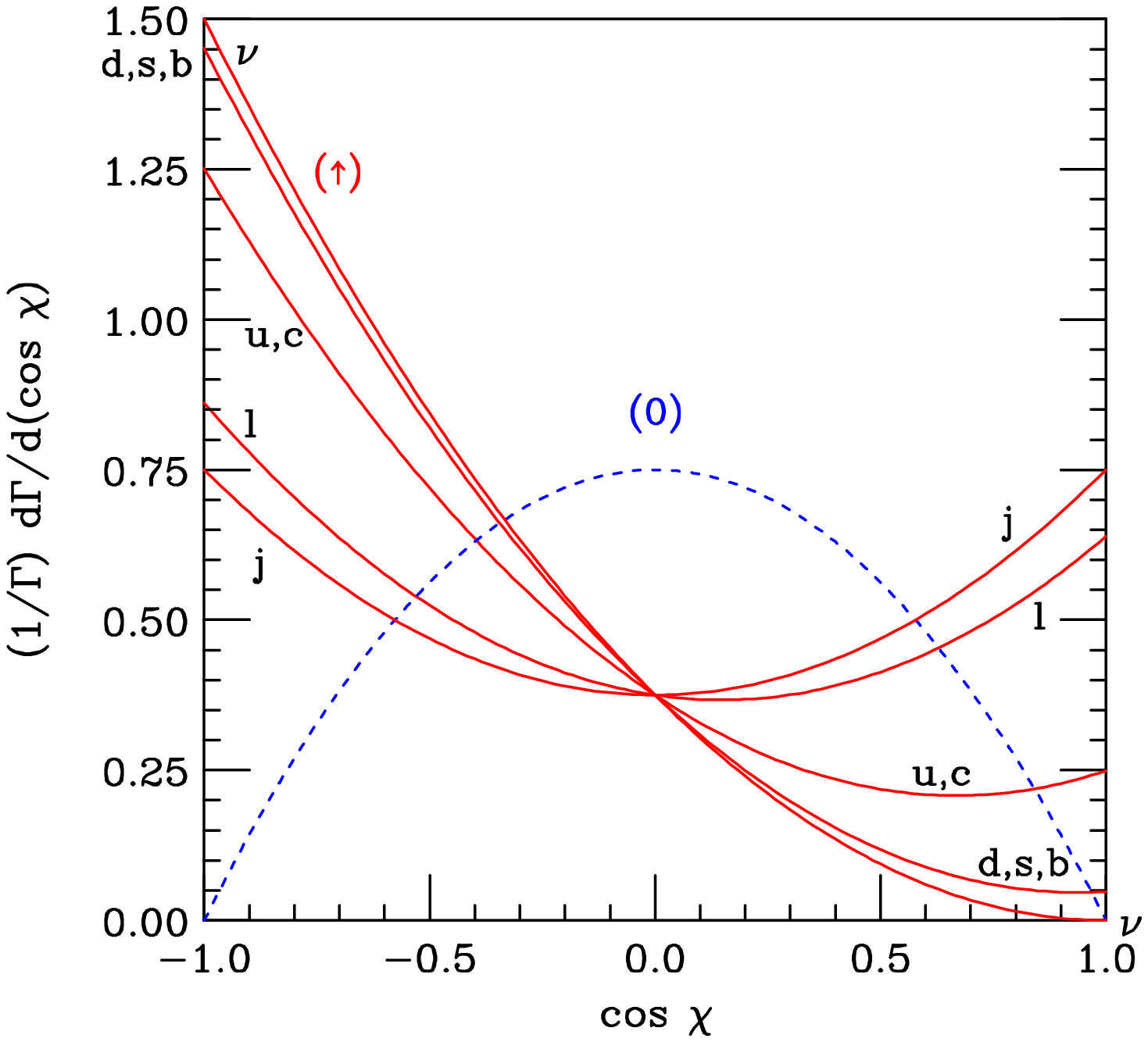}
\vspace{0.0cm}
\caption[]{Angular distributions for the decay of a polarized
$Z$, spin state $\a$ (solid curves) or $(0)$ (dashed curve).  
$\chi$ is the angle between 
the direction of the 
fermion ($\nu, l,u,d,s,c$ or $b$) and the chosen spin axis, 
as viewed in the $Z$ rest frame.  
The decay distribution
for a transversely-polarized $Z$ boson depends
on the flavor of the decay product as labeled; 
the decay of a longitudinally-polarized
$Z$ does not.
$l$ is a generic charged
lepton and may be any of $e^{-}$, $\mu^{-}$, or $\tau^{-}$;
$j$ is a quark or antiquark jet of any flavor.
The distributions for the decay of a $Z$ in the $\b$ spin
state may be generated by replacing $\cos\chi\longrightarrow-\cos\chi$
in the $\a$ distribution.
}
\label{ZDECAYDIST}
\end{figure*}

In these expressions, 
the decay of the $Z$ boson in spin state $\lambda$
to $f_L\bar{f}_R$ is described by the amplitudes
\beqa
\DEE_{L}^{\pm} &\equiv& e^{\pm i\varphi}(\cchi \pm 1),
\cr
\DEE_{L}^{0} &\equiv& \schi\sqrt{2} ,
\label{DecayFactors-L}
\eeqa
whereas the decay to a $f_R\bar{f}_L$ final state depends on
\beqa
\DEE_{R}^{\pm} & = & \DEE_{L}^{\mp*},
\cr
\DEE_{R}^{0} & = & \DEE_{L}^{0*}.
\label{DecayFactors-R}
\eeqa
These distributions have been normalized
to unit area by inclusion of the partial width $\Gamma_f$
for the decay $Z\rightarrow f\bar{f}$.  We define
$\chi$ to be the angle between the direction of motion of the
fermion and the spin axis as seen in the $Z$ rest frame. 
The distributions represented by Eq.~(\ref{Zdecays}) can also
be written
in the form
\beqa
{ {1}\over{\Gamma_f} }
{ {d\Gamma^{\pm}} \over {d(\cos\chi)} }
= {3\over 8}\Biggl[ (1+\cos^2\chi )
\pm 2 \ts
{  
{q_{fL}^2 - q_{fR}^2 }   
\over
{q_{fL}^2 + q_{fR}^2 }   
}
\cos\chi \Biggr]
\label{ZT}
\eeqa
for the transverse polarizations and
\beq
{ {1}\over{\Gamma_f} }
{ {d\Gamma^{0}} \over {d(\cos\chi)} }
= {3\over 4} \sin^2\chi
\label{ZL}
\eeq
for the longitudinal state.
For convenience, we have collected the Standard Model values
of the combination of coupling constants appearing in the $\cos\chi$
term of Eq.~(\ref{ZT}) in Table~\ref{ZDecayTable}.
\begin{table*}
\caption{Coefficients for $Z$ decay.
The numerical values correspond to $\SSQW=0.2312$.
The final entry is for a jet of undetermined charge.
\label{ZDecayTable}}
\begin{ruledtabular}
\begin{tabular}{cccccdcc}
&& fermion &   
$\displaystyle{{q_{fL}^2-q_{fR}^2}\over{q_{fL}^2+q_{fR}^2}}$
&&   \hbox{value} && \\[0.11in]
\colrule
&& $u,c$ &    
   $ \displaystyle{ {9-24\SSQW}\over{9-24\SSQW+32\SSQWSQ} } $
   &=& 0.6686 && \\[0.13in]
&& $d,s,b$ &  
   $ \displaystyle{ {9-12\SSQW}\over{9-12\SSQW+8\SSQWSQ} } $
   &=& 0.9387 && \\[0.13in]
&& $e,\mu,\tau$ &
   $ \displaystyle{ {1-4\SSQW}\over{1-4\SSQW+8\SSQWSQ} } $
   &=&  0.1496 && \\[0.13in]
&& $\nu$ &       
   $1$    
   &=& 1.0000 && \\[0.13in]
&& $q/\bar{q}$ (unspecified) &       
   $0$    
   &=& 0.0000 && \\
\end{tabular}
\end{ruledtabular}
\end{table*}
The corresponding
distributions 
are plotted in Fig.~\ref{ZDECAYDIST}.
Unfortunately, the decays with the most distinctive distributions,
$Z\rightarrow \nu\bar\nu$, are invisible.
On the other hand, for the decays for which charge/flavor identification
would be the easiest,
$Z\rightarrow \ell\bar\ell$, we have a fairly
large overlap between the $\a$ and $\b$ distributions.
Although the charge and flavor identification for
decays to light quarks is virtually impossible,
there is some efficiency for distinguishing $b$ from $\bar{b}$.
Therefore, we will highlight this $Z$ decay mode
when illustrating our results.
%
Even if it turns out to be too optimistic to distinguish
$b$-jets from $\bar{b}$-jets, it will still be possible to
do $Z\HIGGS$/$ZA$ differentiation
because, as we will see shortly,
simply separating the
longitudinal and transverse polarization states 
should be sufficient to provide interesting
information about the nature of the associated Higgs boson.

\vfill\eject\section{Production and Decay Combined}\label{ProdDec}

\subsection{$Zh$ case}

The complete cross section for the
production and decay of a $Z$ and a Higgs may be 
written and understood in terms of the 
cross section
for the production of a $Z$ boson with spin $\lambda$
[Eqs.~(\ref{ZHsigmaLR}) and~(\ref{ZHsigmaRL})]
multiplied by amplitudes describing its decay
[Eqs.~(\ref{DecayFactors-L}) and~(\ref{DecayFactors-R})].
We define the following angles to describe the process:
$\theta^*$ refers to the $Z$ boson production angle
as seen in the ZMF;
$\chi$ is the angle between
the direction of motion of the fermion ({\it i.e.}\ the spatial
part of the 4-vector $f$) and the spin axis as
seen in the $Z$ rest frame;
and $\varphi$ is the azimuthal angle
associated with this decay ({\it i.e.}\ the angle between
the $Z\HIGGS$ production plane and the $Z$ decay plane), also
viewed in the $Z$ rest frame.
With these definitions, the triply-differential cross sections
for $e^{+}e^{-}\longrightarrow Z\HIGGS \longrightarrow f\bar{f} \HIGGS$
read
\beqa
{
{ d^3\sigma_L }
\over
{ d(\cos\theta^{*}) d(\cos\chi) d\varphi}
} &=& \displaystyle{
{3q_{eL}^2}
\over
{ 1024 \pi^3  }
} 
{ 
{ M_{{}_Z}^3 }
  \over
{s \Gamma_{{}_Z} }
} G_F^2 M_W^2 
\Theta(s,M_{\HIGGS},M_{{}_Z})
\cr &\times&
\Biggl\{
q_{fL}^2
\Bigl\vert
\DEE_L^{+} \T_L^{+} + \DEE_L^0 \T_L^0 
+ \DEE_L^{-} \T_L^{-} \Bigr\vert^2
\cr && \enspace +
q_{fR}^2
\Bigl\vert
\DEE_R^{+} \T_L^{+} + \DEE_R^0 \T_L^0 
+ \DEE_R^{-} \T_L^{-} \Bigr\vert^2\Biggr\}
\label{WZ77x-L}
\eeqa
and
\beqa
{
{ d^3\sigma_R }
\over
{ d(\cos\theta^{*}) d(\cos\chi) d\varphi}
} &=& \displaystyle{
{3q_{eR}^2}
\over
{ 1024 \pi^3  }
} 
{ 
{ M_{{}_Z}^3 }
  \over
{s \Gamma_{{}_Z} }
} G_F^2 M_W^2 
\Theta(s,M_{\HIGGS},M_{{}_Z})
\cr &\times&
\Biggl\{
q_{fL}^2
\Bigl\vert
\DEE_L^{+} \T_R^{+} + \DEE_L^0 \T_R^0 
+ \DEE_L^{-} \T_R^{-} \Bigr\vert^2
\cr && \enspace +
q_{fR}^2
\Bigl\vert
\DEE_R^{+} \T_R^{+} + \DEE_R^0 \T_R^0 
+ \DEE_R^{-} \T_R^{-} \Bigr\vert^2\Biggr\}.
\label{WZ77x-R}
\eeqa

\begin{figure*}
\vspace*{9.0cm}
\includegraphics{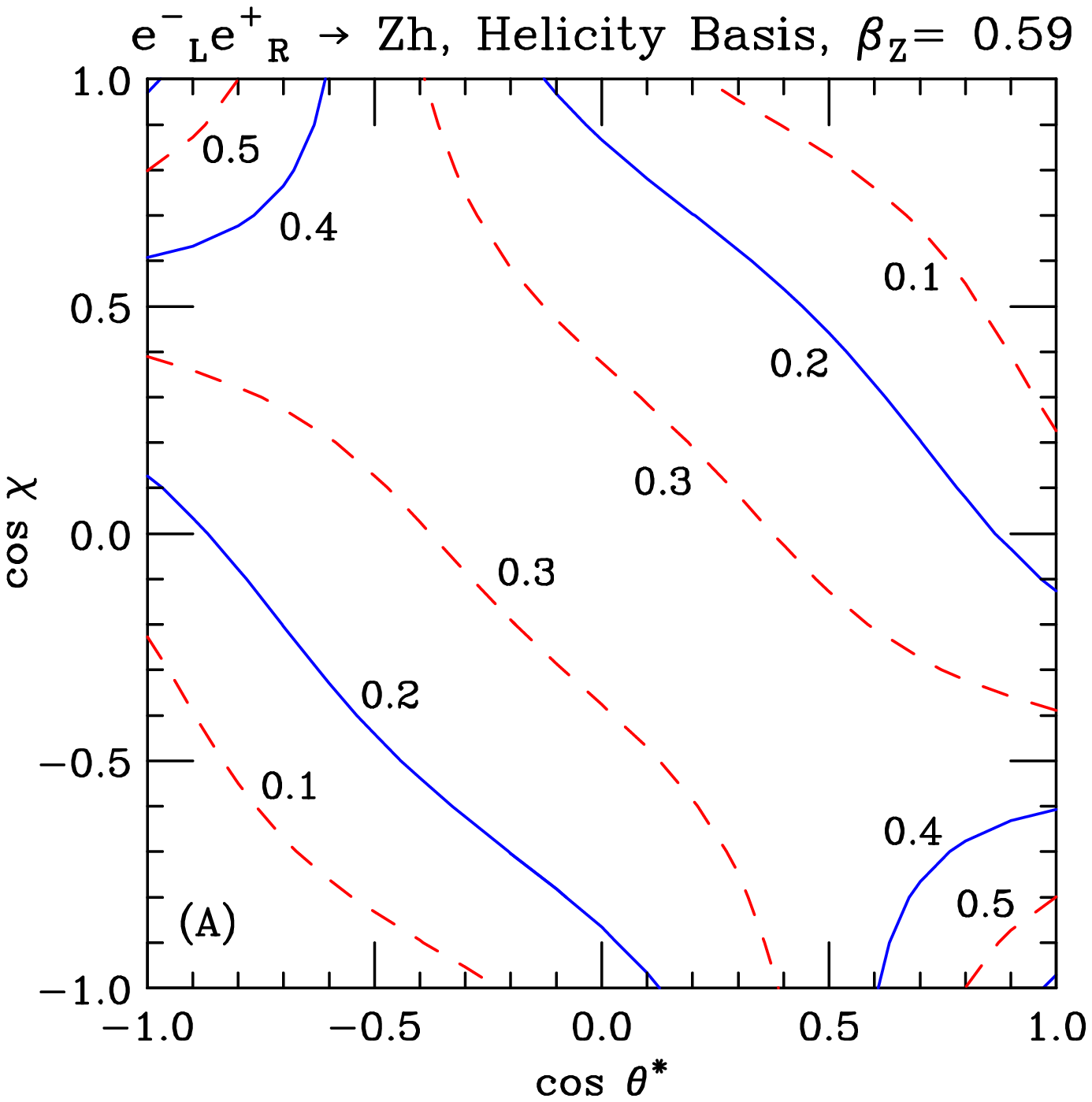}
\includegraphics{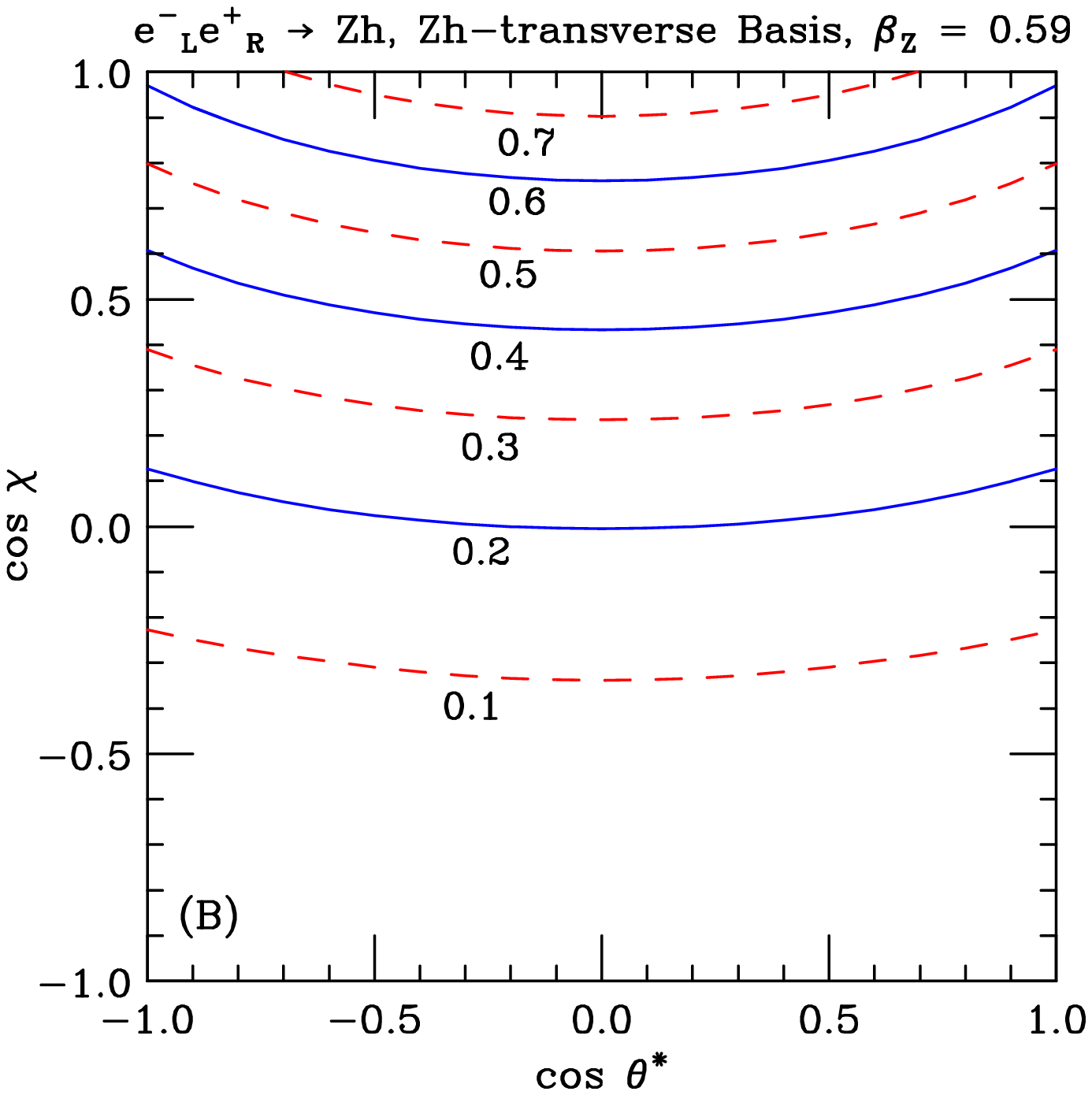}
\caption[]{Double differential production and decay distributions
for $\eebar\rightarrow Z\HIGGS \rightarrow b\bar{b} \HIGGS$,
normalized to unity,
assuming $\sqrt{s} = 250$ GeV and $M_h = 120$ GeV
($\beta_{{}_{Z}}=0.59$).   
Shown are the results using {\bf (A)}\ the helicity basis and
{\bf (B)}\ the $Z\HIGGS$-transverse basis.
}
\label{d2sigmaZH}
\end{figure*}

Since the gamut of potentially interesting spin bases does not
extend to those with dependence on the azimuthal angle, we may
deal with $\varphi$ once and for all by inserting the expressions
for the decay amplitudes contained 
in Eqs.~(\ref{DecayFactors-L}) and~(\ref{DecayFactors-R})
into Eq.~(\ref{WZ77x-L}), 
performing the azimuthal integration, and doing a bit of rearrangement
to arrive at
\beqa
{
{ d^2\sigma_L }
\over
{ d\cth d\cchi }
} & = &
{ { 3q_{eL}^2 }
  \over
  {512 \pi^3}
}
(q_{fL}^2 + q_{fR}^2)
{
{M_{{}_Z}^3}
\over
{s\Gamma_{{}_Z}}
}
G_F^2 M_W^2 
\Theta(s,M_{\HIGGS}, M_{{}_Z} )
\cr && \qquad\times
\Biggl\{
(1+\cchi^2) \vert\Em^+\vert^2
+ 2\schi^2 \vert\Em^0\vert^2
+ (1+\cchi^2) \vert\Em^-\vert^2
\cr  && \qquad\quad\quad +
{
{ q_{fL}^2 - q_{fR}^2 }
\over
{ q_{fL}^2 + q_{fR}^2 }
}
(2 \cos\chi)
\Bigl[
\vert\Em^+\vert^2
-\vert\Em^-\vert^2
\Bigr]
\Biggr\},
\label{WZ790F-L}
\eeqa
where we have defined
\beq
\Em^{+} \equiv \T_L^{+} = \T_R^{-}; \quad
\Em^{-} \equiv \T_L^{-} = \T_R^{+}; \quad
\Em^{0} \equiv \T_L^{0} = \T_R^{0}.
\eeq
Substitution of the explicit forms of the spin functions
in the $Z\HIGGS$-transverse basis
from Eq.~(\ref{ZH-ZHtrans}) allows us to write
the cross section in 
Eq.~(\ref{WZ790F-L}) in the relatively simple form
\beqa
{
{ d^2\sigma_L }
\over
{ d\cth d\cchi }
} 
\Biggr\vert_{\rm trans}
& = &
{ { 3q_{eL}^2 }
  \over
  {512 \pi^3}
}
(q_{fL}^2 + q_{fR}^2)
{
{M_{{}_Z}^3}
\over
{s\Gamma_{{}_Z}}
}
G_F^2 M_W^2 
\Theta(s,M_{\HIGGS}, M_{{}_Z} )
\cr && \qquad\times
\Biggl\{
(1+\cchi^2) [2-\BETA^2(1+\cth^2)]
\cr  && \qquad\quad\quad +
{
{ q_{fL}^2 - q_{fR}^2 }
\over
{ q_{fL}^2 + q_{fR}^2 }
}
(2 \cos\chi)
\Bigl[
2\ginv\sqrt{1-\BETA^2\cth^2}\ts
\Bigr]
\Biggr\}.
\label{WZ834}
\eeqa
Using the 
helicity basis instead leads to a
slightly more complicated result:
\beqa
{
{ d^2\sigma_L }
\over
{ d\cth d\cchi }
} 
\Biggr\vert_{\rm helicity}
& = &
{ { 3q_{eL}^2 }
  \over
  {512 \pi^3}
}
(q_{fL}^2 + q_{fR}^2)
{
{M_{{}_Z}^3}
\over
{s\Gamma_{{}_Z}}
}
G_F^2 M_W^2 
\Theta(s,M_{\HIGGS}, M_{{}_Z} )
\cr && \qquad\times
\Biggl\{
(1-\BETA^2)
(1+\cchi^2) 
(1+\cth^2) + 2 \schi^2\sth^2
\cr  && \qquad\quad\quad - 4 \GAMMA^{-1}\ts
{
{ q_{fL}^2 - q_{fR}^2 }
\over
{ q_{fL}^2 + q_{fR}^2 }
}
\cos\chi\cos\theta^*
\Biggr\}.
\label{WZ837}
\eeqa
The distributions contained in Eqs.~(\ref{WZ834})
and~(\ref{WZ837}) have been plotted in Fig.~\ref{d2sigmaZH}
at $\BETA=0.59$: that is, for a Higgs mass of 120 GeV
and a collider center-of-mass energy of 250 GeV.

The analogous procedure applied to Eq.~(\ref{WZ77x-R})
leads to a similar result which may be generated from
Eq.~(\ref{WZ790F-L}) by the 
interchanges $q_{eL} \leftrightarrow q_{eR}$;
$q_{fL} \leftrightarrow q_{fR}$.
Performing the average over the initial spins leads to the unpolarized
result
\beqa
{
{ d^2\sigma_U }
\over
{ d\cth d\cchi }
} & = &
{ {3}
  \over
  {2048 \pi^3}
}
(q_{eL}^2 + q_{eR}^2)(q_{fL}^2 + q_{fR}^2)
{
{ M_{{}_Z}^3 }
\over
{ s\Gamma_{{}_Z} }
}
G_F^2 M_W^2 
\Theta(s,M_{\HIGGS}, M_{{}_Z} )
\cr && \qquad\times
\Biggl\{
(1+\cchi^2) \vert\Em^+\vert^2
+ 2\schi^2 \vert\Em^0\vert^2
+ (1+\cchi^2) \vert\Em^-\vert^2
\cr  && \qquad\quad\quad + 
{
{ q_{eL}^2 - q_{eR}^2 }
\over
{ q_{eL}^2 + q_{eR}^2 }
}
{
{ q_{fL}^2 - q_{fR}^2 }
\over
{ q_{fL}^2 + q_{fR}^2 }
}
(2 
\cos\chi)
\Bigl[
\vert\Em^+\vert^2
-\vert\Em^-\vert^2
\Bigr] \Biggr\}.
\label{WZ790F-U}
\eeqa

The differences in the shapes of the polarized and unpolarized
distributions are completely contained in the forward-backward
asymmetric ($\cos\chi$)
term, the coefficient of which depends on the difference between
the left and right hand fermion-to-$Z$ couplings as well as the
flavor of fermion involved.  
It should be clear from this discussion plus
the similarities between Eqs.~(\ref{WZ790F-L})
and~(\ref{WZ790F-U}) that it is not necessary
to perform the full calculation separately for each of the three cases:
knowledge of just one case plus
suitable alteration of the coefficient of the $\cos\chi$ term
is sufficient to generate the other two distributions.
For polarized beams the asymmetry factor depends only on the
nature of the $Z$ decay products, whereas if the beams are unpolarized,
an additional factor involving the electron-to-$Z$ couplings dilutes
the forward-backward asymmetry.  
A glance at Table~\ref{ZDecayTable} reveals that this dilution
factor is rather small, only about 0.15; therefore,
the observation of forward-backward asymmetries will be 
greatly aided
by the use of polarized beams.

By choosing a slightly different rearrangement/grouping of the
terms in Eq.~(\ref{WZ790F-L}) and incorporating the information
on the shape of the polarized $Z$ decay distributions
contained in Eqs.~(\ref{ZT}) and~(\ref{ZL}), we conclude
that the fraction of $Z$'s produced in spin state $\lambda$
from $e^{-}_L e^{+}_R$ scattering
may be calculated from
\beq
f^\lambda_L =
{
{ \displaystyle\int_{-1}^{1} d\cth \vert\T_L^\lambda\vert^2 }
\over
{ \displaystyle\int_{-1}^{1} d\cth \Bigl\{ \vert\T_L^{+}\vert^2 
+\vert\T_L^{0}\vert^2 +\vert\T_L^{-}\vert^2 \Bigr\} }
}.
\label{fractanal}
\eeq
For $e_R^- e_L^+$ scattering, we should replace the $\T_L$'s
with $\T_R$'s in the above formula; for unpolarized beams, we should
form the appropriate weighted
average of the left-handed and right-handed fractions:
\beq
f_U^\lambda = 
{ {q_{eL}^2}\over{ q_{eL}^2 + q_{eR}^2 } } f_L^\lambda
+ { {q_{eR}^2}\over{ q_{eL}^2 + q_{eR}^2 } } f_R^\lambda 
\label{fractnonpolar}
\eeq
Eqs.~(\ref{fractanal}) and~(\ref{fractnonpolar}) may be used to determine
the spin fractions using any spin basis 
in any model for which the polarized production
amplitudes $\T_{L,R}^\lambda$ have been calculated.  

\subsection{$ZA$ case}

\begin{figure*}
\vspace*{9.0cm}
\includegraphics{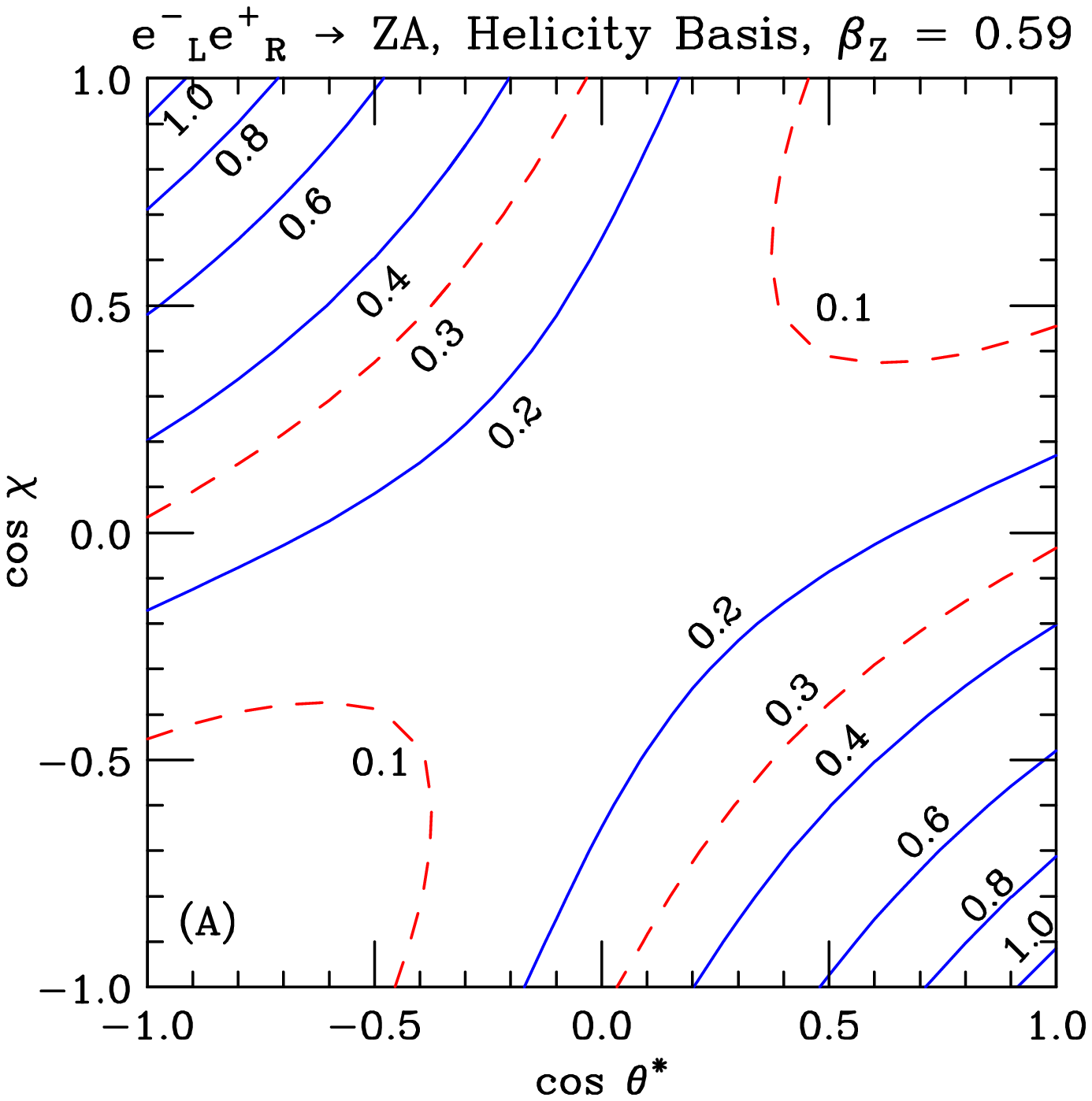}
\includegraphics{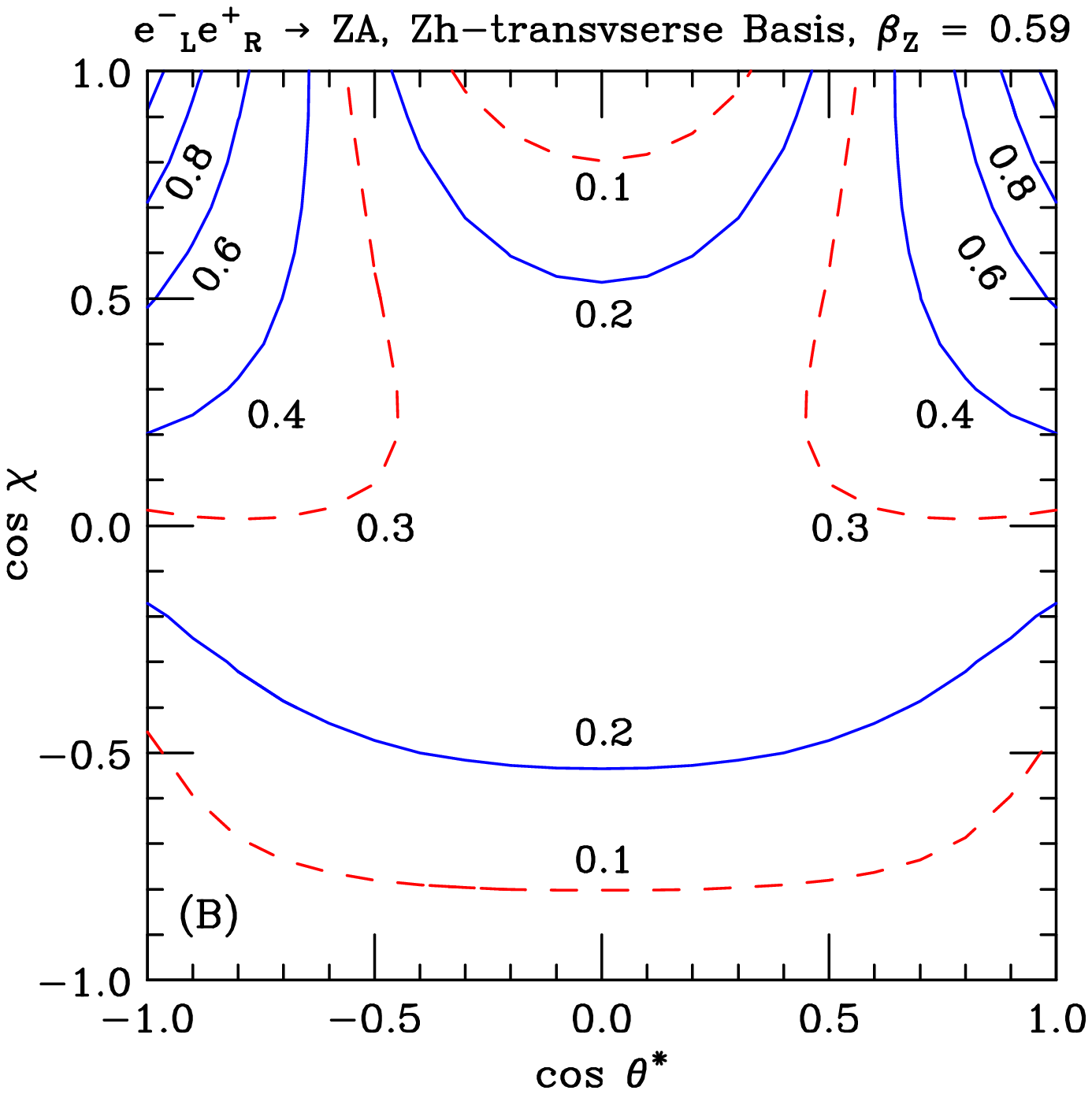}
\caption[]{Double differential production and decay distributions
for $\eebar\rightarrow ZA \rightarrow b\bar{b} A$,
normalized to unity,
assuming $\sqrt{s} = 250$ GeV and $M_h = 120$ GeV
($\beta_{{}_{Z}}=0.59$).   
Shown are the results using {\bf (A)}\ the helicity basis and
{\bf (B)}\ the $Z\HIGGS$-transverse basis.
}
\label{d2sigmaZA}
\end{figure*}

We now turn to the analogous treatment for the case of a 
$CP$-odd boson.  We begin by writing the
the triply-differential cross-section for
the production and decay 
process $e_L^- e_R^+ \rightarrow ZA \rightarrow f\bar{f} A$
in terms of the previously-defined production 
[Eqs.~(\ref{ZA-LR}) and~(\ref{ZA-RL})]
and decay 
[Eqs.~(\ref{DecayFactors-L}) and~(\ref{DecayFactors-R})]
amplitudes:
\beqa
{
{ d^3\sigma_L }
\over
{ d(\cos\theta^{*}) d(\cos\chi) d\varphi}
} &=& \displaystyle{
{ 3 \BETA^2 }
\over
{ 128 \pi^3  }
} 
{
{ M_{{}_Z} }
\over
{ \Gamma_{{}_Z} }
}
\eta^2  {{M_{{}_Z}^4}\over{\Lambda^4}} 
{ {q_{eL}^2}
} G_F^2 M_W^2 
\Theta(s,M_{{}_A},M_{{}_Z})
\cr &\times&
\Biggl\{
q_{fL}^2
\Bigl\vert
\DEE_L^{+} \widetilde\T_L^{+} + \DEE_L^0 \widetilde\T_L^0 
+ \DEE_L^{-} \widetilde\T_L^{-} \Bigr\vert^2
\cr && \enspace +
q_{fR}^2
\Bigl\vert
\DEE_R^{+} \widetilde\T_L^{+} + \DEE_R^0 \widetilde\T_L^0 
+ \DEE_R^{-} \widetilde\T_L^{-} \Bigr\vert^2\Biggr\}.
\label{ZA-WZ77x-L}
\eeqa
As in the $Z\HIGGS$ case, we are not interested in spin bases
that explicitly depend on $\varphi$.  Thus, we 
integrate over $\varphi$ by inserting the explicit decay amplitudes
for the $Z$ boson and perform a bit of algebra to obtain
\beqa
{
{ d^2\sigma_L }
\over
{ d\cth d\cchi }
} & = &
{ { 3\BETA^2 }
  \over
  { 64 \pi^2}
}
{
{ M_{{}_Z} }
\over
{ \Gamma_{{}_Z} }
}
\eta^2  {{M_{{}_Z}^4}\over{\Lambda^4}} 
\ts q_{eL}^2 (q_{fL}^2 + q_{fR}^2)
G_F^2 M_W^2 
\Theta(s,M_{{}_A}, M_{{}_Z} )
\cr && \qquad\quad\times
\Biggl\{
(1+\cchi^2) \vert\widetilde\Em^+\vert^2
+ 2\schi^2 \vert\widetilde\Em^0\vert^2
+ (1+\cchi^2) \vert\widetilde\Em^-\vert^2
\cr  && \qquad\quad\quad\quad +
{
{ q_{fL}^2 - q_{fR}^2 }
\over
{ q_{fL}^2 + q_{fR}^2 }
}
(2 \cos\chi)
\Bigl[
\vert\widetilde\Em^+\vert^2
-\vert\widetilde\Em^-\vert^2
\Bigr]
\Biggr\},
\label{ZA-WZ790F-L}
\eeqa
where it is natural to introduce the following definitions:
\beq
\widetilde\Em^{+} \equiv \widetilde\T_L^{+} = \widetilde\T_R^{-}; \quad
\widetilde\Em^{-} \equiv \widetilde\T_L^{-} = \widetilde\T_R^{+}; \quad
\widetilde\Em^{0} \equiv \widetilde\T_L^{0} = \widetilde\T_R^{0}.
\eeq
The explicit result for the cross section contained in 
Eq.~(\ref{ZA-WZ790F-L}) is reasonably simple in the helicity basis:
\beqa
{
{ d^2\sigma_L }
\over
{ d\cth d\cchi }
} \Biggr\vert_{\rm helicity}
& = &
{ { 3\BETA^2 }
  \over
  { 64 \pi^2}
}
{
{ M_{{}_Z} }
\over
{ \Gamma_{{}_Z} }
}
\eta^2  {{M_{{}_Z}^4}\over{\Lambda^4}} 
\ts q_{eL}^2 (q_{fL}^2 + q_{fR}^2)
G_F^2 M_W^2 
\Theta(s,M_{{}_A}, M_{{}_Z} )
\cr && \qquad\quad\times
\Biggl\{
(1+\cchi^2) (1+\cth^2)
-4{
{ q_{fL}^2 - q_{fR}^2 }
\over
{ q_{fL}^2 + q_{fR}^2 }
}
 \cos\chi\cos\theta^*
\Biggr\},
\label{WZ836}
\eeqa
On the other hand, using the (``wrong'') $Z\HIGGS$-transverse
basis for the spin functions instead yields
\beqa
{
{ d^2\sigma_L }
\over
{ d\cth d\cchi }
} \Biggr\vert_{\rm trans}
& = &
{ { 3\BETA^2 }
  \over
  { 64 \pi^2}
}
{
{ M_{{}_Z} }
\over
{ \Gamma_{{}_Z} }
}
\eta^2  {{M_{{}_Z}^4}\over{\Lambda^4}} 
\ts q_{eL}^2 (q_{fL}^2 + q_{fR}^2)
G_F^2 M_W^2 
\Theta(s,M_{{}_A}, M_{{}_Z} )
\cr && \qquad\quad\times
\Biggl\{
(1+\cchi^2) \ts\cth^2\ts 
{
{2-\BETA^2(1+\cth^2)}
\over
{1-\BETA^2\cth^2}
}
+ 2\ts {{\schi^2\sth^2}\over{1-\BETA^2\cth^2}}
\cr  && \qquad\quad\quad\quad +
{
{ q_{fL}^2 - q_{fR}^2 }
\over
{ q_{fL}^2 + q_{fR}^2 }
}
(4 \cos\chi\cos^2\theta^*)
{ {\sqrt{1-\BETA^2}}
  \over
  {\sqrt{1-\BETA^2\cth^2}}
}
\Biggr\},
\label{WZ839}
\eeqa
These two distributions are compared in Fig.~\ref{d2sigmaZA}.

The result of the corresponding calculation for 
the $e^{-}_R e^{+}_L$ initial
state may be obtained from Eq.~(\ref{ZA-WZ790F-L}) by the
interchanges $q_{eL}\leftrightarrow q_{eR}$; 
$q_{fL}\leftrightarrow q_{fR}$.
Combining the two beam polarizations 
in the appropriate manner leads to the unpolarized production
and decay distribution
\beqa
{
{ d^2\sigma_U }
\over
{ d\cth d\cchi }
} & = &
{ {3 \BETA^2 }
  \over
  {256 \pi^2}
}
{
{M_{{}_Z}}
\over
{\Gamma_{{}_Z}}
}
\eta^2 
{{M_{{}_Z}^4}\over{\Lambda^4}} 
(q_{eL}^2 + q_{eR}^2)(q_{fL}^2 + q_{fR}^2)
G_F^2 M_W^2 
\Theta(s,M_{{}_A}, M_{{}_Z} )
\cr && \quad\qquad\times
\Biggl\{
(1+\cchi^2) \vert\widetilde\Em^+\vert^2
+ 2\schi^2 \vert\widetilde\Em^0\vert^2
+ (1+\cchi^2) \vert\widetilde\Em^-\vert^2
\cr  && \quad\qquad\quad\quad + 
{
{ q_{eL}^2 - q_{eR}^2 }
\over
{ q_{eL}^2 + q_{eR}^2 }
}
{
{ q_{fL}^2 - q_{fR}^2 }
\over
{ q_{fL}^2 + q_{fR}^2 }
}
(2 
\cos\chi)
\Bigl[
\vert\widetilde\Em^+\vert^2
-\vert\widetilde\Em^-\vert^2
\Bigr] \Biggr\}.
\label{ZA-WZ790F-U}
\eeqa
A careful comparison of Eqs.~(\ref{ZA-WZ790F-L})
and~(\ref{ZA-WZ790F-U}) reveals that the unpolarized distribution
may be obtained from the left-handed distribution by replacing
$q_{eL}^2 \rightarrow {1\over4}(q_{eL}^2+q_{eR}^2)$ in the 
prefactor and adjusting the coefficient of the term linear
in $\cos\chi$ as follows:
\beq
{
{ q_{fL}^2 - q_{fR}^2 }
\over
{ q_{fL}^2 + q_{fR}^2 }
} \rightarrow
{
{ q_{eL}^2 - q_{eR}^2 }
\over
{ q_{eL}^2 + q_{eR}^2 }
} \ts
{
{ q_{fL}^2 - q_{fR}^2 }
\over
{ q_{fL}^2 + q_{fR}^2 }
} .
\eeq

The fraction of $Z$ bosons produced in each of the three
possible spin states may be calculated from an expression
similar to Eqn.~(\ref{fractanal}), but with the ${\T}$'s
replaced by $\widetilde{\T}$'s.

\vfill\eject\section{$Z$ decay angular distributions}\label{enchiladas}
At last we come to the decay angular distribution $d\sigma/d(\cos\chi)$
which will be useful in distinguishing $Z\HIGGS$ from $ZA$.
Since the choice of spin basis influences the exact definition
of $\chi$, this distribution will depend on the choice made for $\xi$.
Put differently, the decay angular distribution may be written
in the form
\beq
{{1}\over{\sigma}}\ts
{{d\sigma}\over{d(\cos\chi)}}
= f^+ \ts {{1}\over{\Gamma_f}}
{{d\Gamma^+}\over{d(\cos\chi)}}
+ f^0 \ts {{1}\over{\Gamma_f}}
{{d\Gamma^0}\over{d(\cos\chi)}}
+ f^- \ts {{1}\over{\Gamma_f}}
{{d\Gamma^-}\over{d(\cos\chi)}},
\label{FractionalEnchiladas}
\eeq
where the spin fractions may be calculated from Eq.~(\ref{fractanal})
for polarized beams or from Eq.~(\ref{fractnonpolar}) for unpolarized
beams. The unit-normalized $Z$ decay distributions are given
by Eqs.~(\ref{ZT}) and~(\ref{ZL}).
That is, the decay angular distribution is simply a
linear superposition of the polarized $Z$ decay distributions weighted
by the fraction of $Z$'s in each spin state.
Clearly the fraction of $Z$'s with
a given spin depends on the choice of spin axis.
The decay distribution $d\sigma/d(\cos\chi)$ in a particular spin
basis may be calculated by 
inserting the desired expressions for $\sxi$ and $\cxi$ 
into the spin functions~(\ref{ZH-LR})
and~(\ref{ZH-RL}), and then using the results to calculate
the spin fractions from
Eq.~(\ref{fractanal}).

\subsection{$Z\HIGGS$ in the helicity basis}
\begin{figure*}

\vspace*{9cm}
\includegraphics{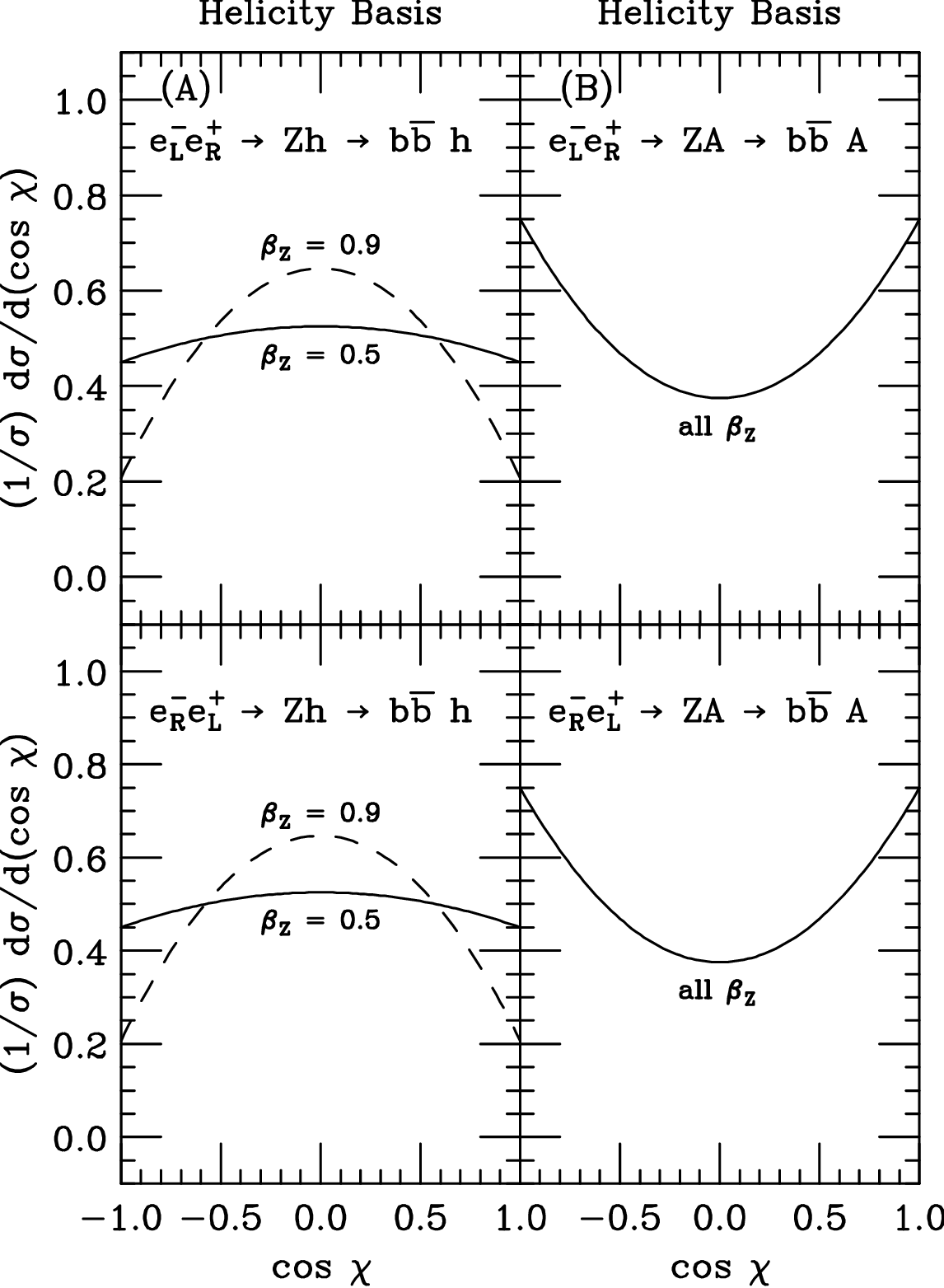}
\vspace{8.5cm}

\caption[]{$Z$ decay angular distributions 
in the ZMF helicity basis for
{\bf (A)}
$\eebar\rightarrow Z\HIGGS\rightarrow b\bar{b} h$ 
and 
{\bf (B)}
$\eebar\rightarrow ZA\rightarrow b\bar{b} A$.
The decay angle 
$\chi$
is defined as the angle in the $Z$ rest frame between 
the spin axis direction and the direction of motion of the 
negatively-charged lepton.
}
\label{dSIGMA-dCCHI-helicity}
\end{figure*}

For example, setting $\xi = \pi$ to
obtain the helicity basis leads to
\beqa
{{1}\over{\sigma_L}}\ts
{{d\sigma_L}\over{d(\cos\chi)}} 
\Biggr\vert_{\rm helicity}=
{{3}\over{12-8 \BETA^2}}
\biggl[ 2 - \BETA^2(1 + \cos^2\chi ) \biggr].
\label{WZ684}
\eeqa
In the panels on the left-hand side of
Fig.~\ref{dSIGMA-dCCHI-helicity} we have plotted 
this distribution for $\BETA=0.5$ and~0.9.  
The distribution is nearly flat in $\cos\chi$ close to threshold
($\bz = 0$), and becomes more and more concave down as 
$\bz$ is increased.
For $\bz \rightarrow 1$ (the ultra-relativistic limit) we have
\beqa
{{1}\over{\sigma_T}}\ts
{{d\sigma}\over{d(\cos\chi)}} 
\Biggr\vert_{\rm helicity} \longrightarrow
{{3}\over{4}} \sin^2 \chi,
\label{HEL-UR}
\eeqa
the signature of $Z$ bosons produced with 100\% longitudinal
polarization in accordance with the vector
boson equivalence theorem.

\subsection{$Z\HIGGS$ in the \ZHtrans\ basis}
\begin{figure*}
\vspace*{9cm}
\includegraphics{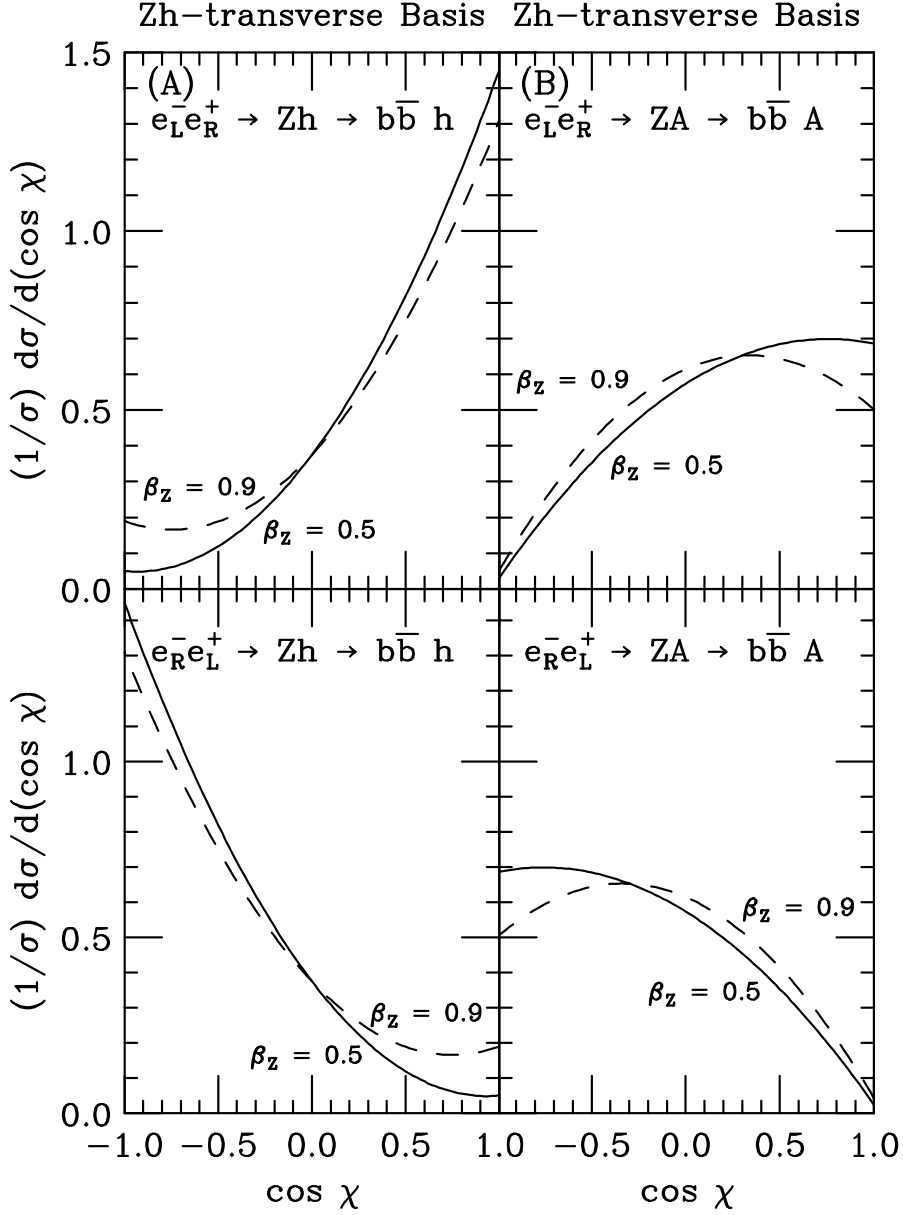}
\vspace{8.5cm}
\caption[]{$Z$ decay angular distributions 
in the \ZHtrans\ basis for
{\bf (A)}~$\eebar\rightarrow Z\HIGGS\rightarrow b \bar{b} h$ 
and {\bf (B)}~$\eebar\rightarrow ZA\rightarrow b \bar{b} A$.
The decay angle 
$\chi$ 
is defined as the angle in the $Z$ rest frame between 
the spin axis direction and the direction of motion of the 
negatively-charged lepton.
}
\label{dSIGMA-dCCHI-trans}
\end{figure*}

We now turn to the \ZHtrans\ basis, Eq.~(\ref{magic-xi}), 
which was engineered
to eliminate the longitudinal $Z$ bosons from the mix.
Consequently, we obtain a decay angular distribution 
of the form
\beqa
{{1}\over{\sigma_L}}\ts
{{d\sigma_L}\over{d(\cos\chi)}}  
\Biggr\vert_{\rm trans} =
{{3}\over{8}} (\cos^2\chi + 1)
+ {{9}\over{8}} 
{ {q_{fL}^2-q_{fR}^2 }
\over
{ q_{fL}^2+q_{fR}^2 } }
{{\cos\chi}\over{3 - 2\BETA^2 }}
\Bigl( 1-\BETA^2 + {{\arcsin\BETA}\over{\GAMMA\BETA}} \Bigr).
\label{WZ689C}
\eeqa
A plot of this distribution appears in the panels on the left side
of Fig.~\ref{dSIGMA-dCCHI-trans}.
Eq.~(\ref{WZ689C}) simplifies near threshold ($\bz\longrightarrow0$)
and in the ultra-relativistic limit ($\bz\longrightarrow1$) as follows:
\beqa
{{1}\over{\sigma_L}}\ts
{{d\sigma_L}\over{d(\cos\chi)}}  
\Biggr\vert_{\rm trans} && \longrightarrow
{{3}\over{8}} (\cos^2\chi + 1)
+ {{3}\over{4}} 
{ {q_{fL}^2-q_{fR}^2 }
\over
{ q_{fL}^2+q_{fR}^2 } }
\cos\chi \bigl[1 + {\cal O}(\bz^4) \bigr]
\quad(\bz \longrightarrow 0)
\cr && \longrightarrow
{{3}\over{8}} (\cos^2\chi + 1) \quad
\qquad\qquad\qquad
\qquad\qquad\qquad\quad\enspace\ts
(\bz \longrightarrow 1).
\label{TRANSlimits}
\eeqa
Notice that in contrast to the result in the helicity basis, 
even near threshold
this distribution displays non-trivial correlations.
The shape of this distribution
is a rather weak function of the machine energy.

\subsection{$ZA$ in the helicity basis}

Repeating the calculation with the pseudoscalar spin functions
instead, we obtain the result for $ZA$ production and
decay in the helicity basis:
\beq
{{1}\over{\sigma_L}}\ts
{{d\sigma_L}\over{d(\cos\chi)}} 
\Biggr\vert_{\rm helicity}
= {{3}\over{8}} (1 + \cos^2 \chi).
\label{WZ739E}
\eeq
This angular distribution is plotted in 
the panels on the right-hand side of
Fig.~\ref{dSIGMA-dCCHI-helicity}.
There is no forward-backward asymmetry in Eq.~(\ref{WZ739E})
due to the equal mix of $\a$ and $\b$ spin states in this basis.
Furthermore, since we have used a fixed value of $\xi$ and
the spin functions contain no $\BETA$ dependence themselves, the
result in Eq.~(\ref{WZ739E}) holds for all machine energies and
all boson masses (provided, of course, that the $ZA$ final state
is kinematically allowed).


\subsection{$ZA$ in the \ZHtrans\ basis}
Since we won't know {\it a priori}\ what sort of Higgs we are dealing
with, we also present the $ZA$ decay angular distribution in the
(``wrong'') Z\HIGGS-transverse basis:
\beqa
{{1}\over{\sigma_L}}\ts
{{d\sigma_L}\over{d(\cos\chi)}} 
\Biggr\vert_{\rm trans} &&
= (1 + \cos^2\chi)
\Biggl[ {{3}\over{32}} - {{9}\over{64}} {{1-\BETA^2}\over{\BETA^2}}
\biggl(  {{1}\over{\BETA}} \ln {{1-\BETA}\over{1+\BETA}}
+ 2 \biggr) \Biggr]
\cr && \quad + {{9}\over{32}} \sin^2\chi
\Biggl[ {{2}\over{\BETA^2}} + 
{{1-\BETA^2}\over{\BETA^3}} \ln {{1-\BETA}\over{1+\BETA}}
\Biggr]
\cr && \quad - {{9}\over{16}} 
{ {q_{fL}^2-q_{fR}^2 }
\over
{ q_{fL}^2+q_{fR}^2 } }
\cos\chi \GINV
{{1}\over{\BETA^2}}
\Biggl[ 1 - \BETA^2 -  {{\arcsin \BETA}\over{\BETA\GAMMA}} 
\Biggr].
\label{WZ739U}
\eeqa
A plot of this angular distribution appears in 
Fig.~\ref{dSIGMA-dCCHI-trans}B.
Eq.~(\ref{WZ739U}) looks somewhat complicated
because it employs a spin basis designed to simplify
the $Z\HIGGS$ amplitude, not the $ZA$ amplitude.
All of the $\BETA$-dependence contained in Eq.~(\ref{WZ739U})
is a consequence of choosing $\xi$ to be an explicit function of $\BETA$
[{\it cf.}\ Eq.~(\ref{magic-xi})].


\subsection{Comparison of $Zh$ and $ZA$}\label{Zh-vs-ZA}

\begin{figure*}
\vspace*{9cm}
\includegraphics{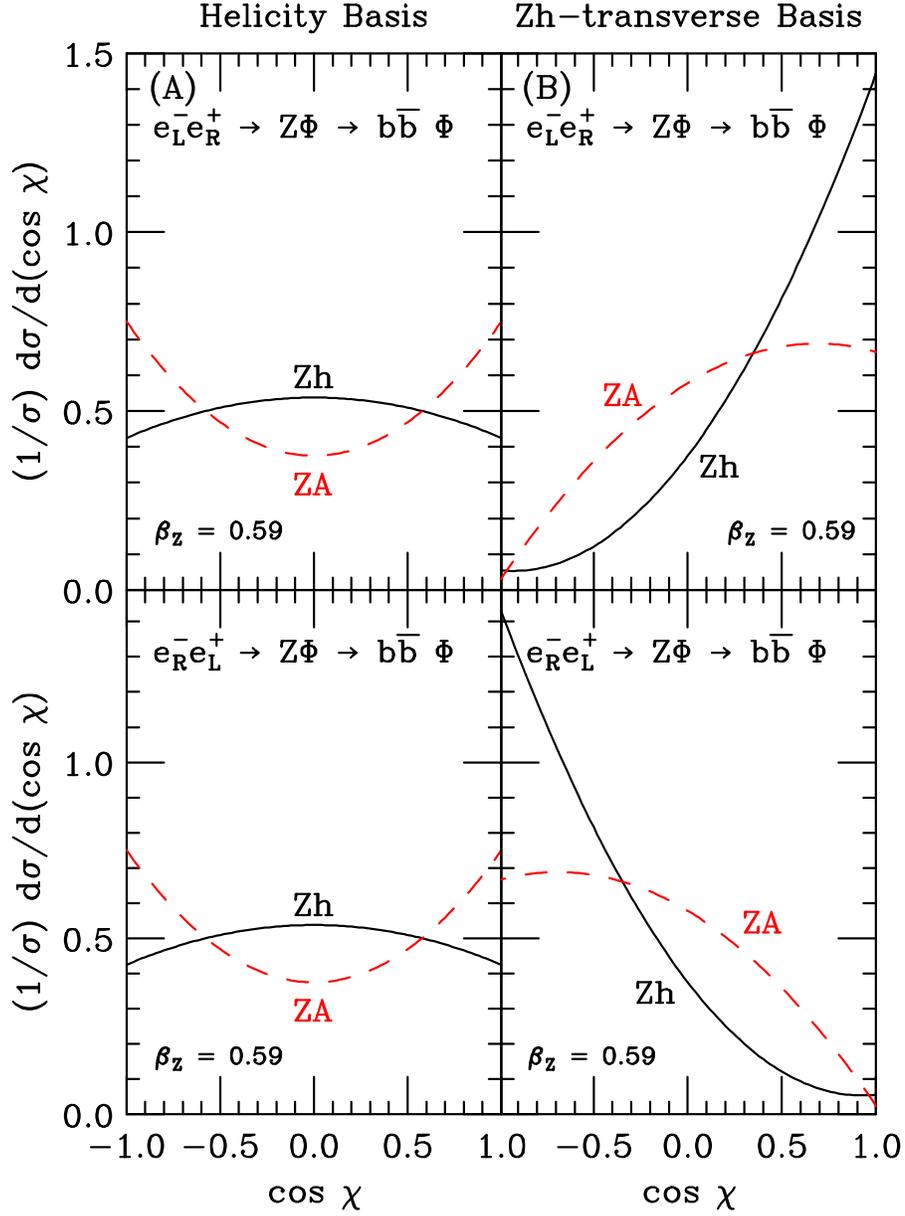}
\vspace{8.5cm}
\caption[]{Comparison of $Z$ decay angular distributions for
{\bf (A)}~$\eebar\rightarrow Zh\rightarrow b \bar{b} h$ 
and
{\bf (B)}~$\eebar\rightarrow ZA\rightarrow b \bar{b}A$ 
for $\BETA=0.59$ ($M_\HIGGS = 120 \GeV$, $\protect\sqrt{s} = 250 \GeV$)
in the helicity and \ZHtrans\ bases.
The decay angle 
$\chi$ 
is defined as the angle in the $Z$ rest frame between 
the spin axis direction and the direction of motion of the 
negatively-charged lepton.
}
\label{TheWholeEnchilada}
\end{figure*}
We now turn to a direct comparison between
the decay angular distributions for $Z\HIGGS$ and $ZA$ in
both the helicity and
\ZHtrans\ bases.
From Fig.~\ref{TheWholeEnchilada}
we see that, when the $Z\HIGGS$-transverse
basis is used, there are regions of the $\cos\chi$
distribution where the expected number of events differs
by as much as a factor of 2.  In contrast, when the helicity
basis is used, the maximum difference in expected number
of events is never that large.

A second feature of this plot which has the potential to be
used in distinguishing $Z\HIGGS$ from $ZA$ is a number we will
call the ``$\cos\chi$ forward-backward asymmetry ratio'', $\ECKS$:
\beq
\ECKS \equiv
{
\displaystyle{   
\int_{0}^{1} 
d(\cos\chi){{1}\over{\sigma_L}} 
{ {d\sigma_L}\over{d(\cos\chi)} }
}
\over
\displaystyle{   
\int_{-1}^{0} 
d(\cos\chi){{1}\over{\sigma_L}} 
{ {d\sigma_L}\over{d(\cos\chi)} }
}
}.
\label{EnchiladaRatio}
\eeq
In plain English, $\ECKS$ is the ratio of the
number of events with $\cos\chi>0$ to the number of events with
$\cos\chi<0$.  We may use the expressions contained in 
Eqs.~(\ref{ZT}), (\ref{ZL}), and (\ref{FractionalEnchiladas})
to rewrite this in terms of the coupling constants for the chosen
$Z$ decay mode and the fractions of spin-$\a$ and spin-$\b$ $Z$ bosons:
\beq
\ECKS =
{
{ 4 + 3 \ts\displaystyle{ {q_{fL}^2 - q_{fR}^2}
                       \over
                       {q_{fL}^2 + q_{fR}^2} 
                     } \ts(f^{+} - f^{-}) }
\over
{ 4 - 3 \ts\displaystyle{ {q_{fL}^2 - q_{fR}^2}
                       \over
                       {q_{fL}^2 + q_{fR}^2} 
                     } \ts(f^{+} - f^{-}) }
}.
\label{EnchiladaRatioSimplified}
\eeq
We see from Eq.~(\ref{EnchiladaRatioSimplified}) that two ingredients 
are necessary to have $\ECKS\ne 1$:  first, we need charge/flavor
identification in the final state of the $Z$'s (see the last line of
Table~\ref{ZDecayTable}).   Second, we need the fraction of spin-$\a$
$Z$ bosons ($f^{+}$) to differ from the fraction of spin-$\b$ $Z$ bosons
($f^{-}$).  We point this out because in the helicity basis, both
$Z\HIGGS$ and $ZA$ production have equal numbers of spin-$\a$ and
spin-$\b$ $Z$ bosons.  Hence
\beq
\ECKS_{\rm helicity}^{Zh} = 1 ; \qquad\qquad
\ECKS_{\rm helicity}^{ZA} = 1,
\label{ECKS-helicity}
\eeq
{\it i.e.}\ in the helicity basis, measuring the
value of $\ECKS$ does not distinguish between $Z\HIGGS$ and $ZA$.
On the other hand, in the $Z\HIGGS$-transverse basis, the
values in Tables~\ref{ZH-Breakdowns} and~\ref{ZA-Breakdowns} tell
us that
\beq
\ECKS_{\rm trans}^{Zh} = 5.54 ; \qquad\qquad
\ECKS_{\rm trans}^{ZA} = 1.95,
\label{ECKS-transverse}
\eeq
{\it i.e.} in the $Z\HIGGS$-transverse basis, $\ECKS$ is a potentially
very useful measure that can distinguish $Z\HIGGS$ from $ZA$.

Because the optimal spin basis for studying $Z\HIGGS$ production
and decay is not the same as the optimal basis for studying
$ZA$ production and decay, it is not clear at this stage 
exactly which
route is the best one to pursue.  (See Appendix~\ref{BML}
for a discussion of the beamline basis.  Depending on 
the machine energy, the beamline basis provides a competitive
alternative to the $Z\HIGGS$-transverse basis and may possess 
smaller systematic uncertainties.) 
Optimization of the method requires a detailed detector simulation.  
Nevertheless, it is clear that a sound strategy would involve
utilizing all possible sources of information about distributions
that differ between the two processes. 
For example, the turn-on of the cross section as the machine energy
is raised above threshold is different for $Z\HIGGS$ and $ZA$
production (Eq.~(\ref{ZHsigmaDIFF-LR}) 
versus Eq.~(\ref{ZAsigmaDIFF-LR}); also Ref.~\cite{Miller}).

A second distinguishing characteristic of the two processes 
is the radically different
$\xi$-dependence of the $Z\HIGGS$ and $ZA$ amplitudes.
A measurement of the $Z$ spin composition of a Higgs
signal for different choices of $\xi$ could be used 
to provide one piece of evidence relating to the correct assignment
of ${\cal J}^{PC}$ quantum numbers to the state.
In particular, it would be useful to 
measure the fraction of longitudinally polarized $Z$'s in a sample
of $Z$-Higgs candidates in using both the helicity and \ZHtrans\ bases.
For the signal events, a scalar Higgs should show 
no longitudinal $Z$'s in the \ZHtrans\ basis, while for a 
pseudoscalar Higgs the longitudinal $Z$ fraction would be 
in the 50\%--75\% range, depending on the machine energy.
Although one should really do a full detector
simulation to be sure (the systematics associated with each basis 
will be different), we believe that
the difference between 50\% and 0 should
be large enough to be observable with only modest 
detector sensitivity.

%


\subsection{$CP$-Violating Higgs Bosons}\label{CPviol}

Once the additional structure necessary to generate a 
$CP$-odd Higgs boson has been introduced into the Lagrangian,
it is a small step to incorporate some level of $CP$-violation
in the scalar 
sector~\cite{CPviolhiggses}.
Rather than attempt an exhaustive survey of all of the
possible mechanisms and models,
we will briefly consider
possibility that the Higgs mass eigenstate is not a $CP$-eigenstate,
and describe how this would affect the angular distributions
we have been discussing in this paper.

In particular,
we imagine that the physical Higgs mass eigenstates are a linear
combination of the $CP$-even and $CP$-odd states:
\beqa
\phi_1 &\equiv&  h \cos\psi + A \sin\psi
\cr
\phi_2 &\equiv&  -h \sin\psi + A \cos\psi,
\label{CPviolHiggs}
\eeqa
where $\psi$ is an effective mixing angle, defined such that
when $\psi=0$, the state $\phi_1$ is purely $CP$-even and $\phi_2$
purely $CP$-odd.

If the only source of $CP$-violation in the Higgs sector is 
through the mixing in Eq.~(\ref{CPviolHiggs}), then it is 
straightforward to see how the effects will turn up in the
$\cos\chi$ (or other) distribution.  That is, 
the result will be a linear combination of the $Zh$ and $ZA$
distributions weighted by coefficients that are sensitive to
the mixing angle $\psi$.  Thus, if we look at a distribution
(such as $d\sigma/d(\cos\chi)$ using the \ZHtrans\ basis)
for which the pure $Zh$ and $ZA$ predictions differ
significantly,
information on the value of $\psi$ can be extracted by a fit to this
distribution.  The advantage of this method is that this 
measurement can be performed without even
looking at how the Higgs decays.  
Since $Z$ decays are well-understood,
any deviations from the 
$CP$-conserving predictions contained in 
Eqs.~(\ref{WZ689C}), (\ref{WZ739U}), and~(\ref{ECKS-transverse})
can be unambiguously attributed to the $CP$-quantum numbers
(or lack thereof) of the ``Higgs''.


\section{Conclusions}\label{CONC}

Once the Higgs boson is discovered, it will be important to
check its properties to see if its 
spin-parity-charge-conjugation
quantum numbers are indeed 
${\cal J}^{PC}=0^{++}$ as predicted
by the Standard Model, or if some other set of values
(for example, ${\cal J}^{PC}=0^{+-}$)
applies.  As with any physics measurement, it is best to have
multiple approaches so that consistency checks may be performed.

In this paper we have discussed distinguishing between the
associated production of a scalar ($0^{++}$) Higgs boson
with a $Z$ boson 
and the associated production of a (so-called) pseudoscalar ($0^{+-}$)
Higgs boson with a $Z$ boson.
As noted in Ref.~\cite{Miller},
the total cross sections for the
two processes have different energy dependences near threshold:
see Eqs.~(\ref{ZHsigmaTOTAL-LR}) and~(\ref{ZAsigmaTOTAL-LR}).
Furthermore, the 
production-angle distribution for the $Z\HIGGS$
process is proportional to $1-\half\BETA^2(1+\cos^2\theta^*)$~\cite{Barger}:
that is, it
is flat near threshold ($\BETA\rightarrow0$), and
develops a $1-\cos^2\theta^*$ shape at high energies 
($\BETA\rightarrow 1$).  On the other hand, in
the $ZA$ case the shape of this distribution is $1+\cos^2\theta^*$,
irrespective of energy.

The primary focus of this paper and 
an additional means of distinguishing $Z\HIGGS$ production from
$ZA$ production is provided by the decay-angular distributions
of the $Z$ boson illustrated in Fig.~\ref{TheWholeEnchilada}.
This approach has two distinct advantages:
both of which stem from looking at the $Z$ rather than the
Higgs.  First, it provides a method that 
does not depend on the existence of a particular Higgs decay mode
with a sufficiently large branching ratio.  In fact, this method
does not require observation of the Higgs decay products at all!
The second main advantage is that $Z$ decays are well-understood.
Thus, examining
the spin state of the $Z$ can provide unambiguous information
about the type of Higgs boson it was produced with.  In this connection,
unless the collider center-of-mass energy is large enough so that
the $Z$'s are ultra-relativistic, it is fruitful to investigate
other choices for the $Z$-boson spin basis besides the traditional
helicity basis.  In particular, the $Z\HIGGS$-transverse basis,
defined in Eq.~(\ref{magic-xi}), is potentially useful since
$Z\HIGGS$ events should contain {\bf no}\ longitudinal $Z$'s
whereas the fraction of longitudinal $Z$'s in $ZA$ events is
over 50\% (see Tables \ref{ZH-Breakdowns} and~\ref{ZA-Breakdowns}).
This measurement benefits from having polarized beams.
In particular, for the combination of
Higgs mass ($M_\HIGGS$=120 GeV) and machine energy
($\sqrt{s}=250$~GeV) highlighted in this paper,
more than 99\% of the $Z$ bosons in the
$e^{-}_L e^{+}_R \rightarrow Z\HIGGS$ process are 
in the $\a$ spin state when the $Z\HIGGS$-transverse spin
basis is used.  This is very close to the ``ideal'' situation
where all of the $Z$'s are produced with the same spin projection
and leads to angular correlations which are very
nearly as large as theoretically possible.  Large angular
correlations are well-suited to defining quantities like the
$\cos\chi$ forward-backward asymmetry ratio, Eq.~(\ref{EnchiladaRatio}),
that are sensitive to the $CP$-eigenvalue of the Higgs boson.
In the helicity basis, this ratio is predicted to be unity for
both $Z\HIGGS$ and $ZA$ production whereas in the $Z\HIGGS$ transverse
basis we predict 
a significant difference in this ratio
between $Z\HIGGS$ and $ZA$ production.


\begin{acknowledgments}

The Fermi National Accelerator 
Laboratory 
is operated by Universities Research Association,
Inc., under contract DE-AC02-76CHO3000 with the U.S. Department
of Energy.
We would like to thank Peter Fisher and Philip Bambade 
for illuminating discussions.
GM would like to thank Mike Doncheski for helpful discussions,
and the FNAL theory group for their kind
hospitality during visits to concentrate on this work.
Additional support for GM for this work was provided through
a Research Development Grant from the Commonwealth College
of the Pennsylvania State University.

\end{acknowledgments}


\newpage
\appendix
\vfill\eject\section{Role of Interference Terms}

Suppose that one could find a spin basis where the $Z$'s are produced
in a single spin state $\lambda_0$.  
Then, the full production and decay distribution could be computed from
\beq
d\sigma \sim \vert {\cal M}_{\lambda_0} {\cal D}_{\lambda_0} \vert^2.
\label{SimpleCase}
\eeq
The interpretations of the pieces of Eq.~(\ref{SimpleCase}) are
obvious:  ${\cal M}_{\lambda_0}$ is the amplitude for producing the
$Z$ with spin $\lambda_0$ in association with the Higgs;
${\cal D}_{\lambda_0}$ is the amplitude describing the decay
of the $Z$ with spin $\lambda_0$.

Of course, no such basis exists (although, as pointed out
above, the $Z\HIGGS$-transverse basis comes very close).
So, what we really have is of the form
\beq
d\sigma \sim \biggl\vert\sum_{\lambda} {\cal M}_{\lambda} 
{\cal D}_{\lambda} \biggr\vert^2.
\label{NotSoSimpleCase}
\eeq
Our (semi-classical) intuition would like us to interpret this
result as the sum over all possible spin states of the probability
to produce the given spin state multiplied by its decay probability.
The difference between Eq.~(\ref{NotSoSimpleCase}) and the expression
representing our intuition is 
\beq
{\cal I} \equiv
\biggl\vert\sum_{\lambda} {\cal M}_{\lambda} 
{\cal D}_{\lambda} \biggr\vert^2
- \sum_{\lambda} \biggl\vert {\cal M}_{\lambda} 
{\cal D}_{\lambda} \biggr\vert^2.
\label{InterferenceTerms}
\eeq
The contributions on the right hand side of
Eq.~(\ref{InterferenceTerms}) are precisely
the quantum interference terms present in the coherent 
sum~(\ref{NotSoSimpleCase}).
When these interference terms are
small, the contribution to the full production and decay distribution
fits our semi-classical intuition.  
On the other hand, when Eq.~({\ref{InterferenceTerms}) is large,
quantum interference effects become important, and our semi-classical
intuition falls short.  

Thus, one way to judge the quality of a particular spin basis is to
examine the relative importance of the interference terms in the
production and decay distribution.  We normalize the interference
sum to the total matrix element on a point-by-point basis,
\beq
\Ihat \equiv
{ {
  \biggl\vert\displaystyle\sum_{\lambda} {\cal M}_{\lambda} 
  {\cal D}_{\lambda} \biggr\vert^2
  - \displaystyle\sum_{\lambda} \biggl\vert {\cal M}_{\lambda} 
  {\cal D}_{\lambda} \biggr\vert^2
}
\over
{  \biggl\vert\displaystyle\sum_{\lambda} {\cal M}_{\lambda} 
  {\cal D}_{\lambda} \biggr\vert^2
} }
\label{IhatDef}
\eeq
and then determine the distribution in $\Ihat$.
For a process containing
$N$ independent intermediate spin configurations, $\Ihat$
can range from $-\infty$ (total destructive interference)
to $(N-1)/N$ (total constructive interference).
Clearly, in the 
simple case represented by Eq.~(\ref{SimpleCase})
where there is no interference,
this distribution will be a delta function centered at $\Ihat=0$.
In the more general situation represented by Eq.~(\ref{NotSoSimpleCase}),
a plot of $d\sigma/d\Ihat$ displays the relative importance
of the various values of $\Ihat$.  

With these considerations in mind, we present Fig.~\ref{ZHZAinf},
\begin{figure*}
\vspace*{7cm}
\includegraphics{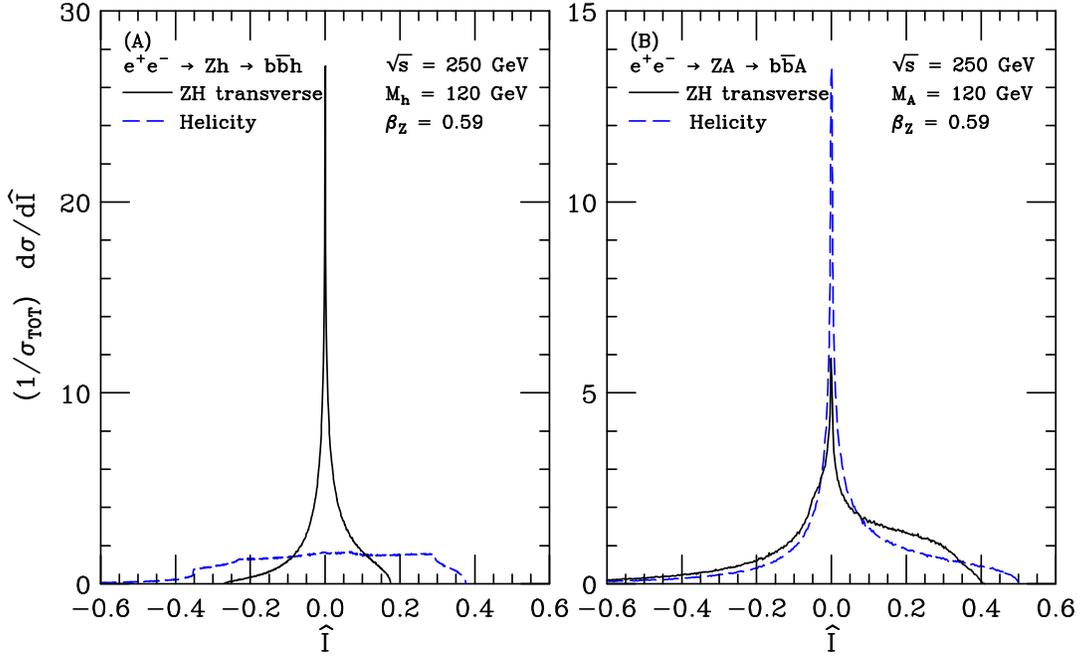}
\vspace{3.5cm}
\caption[]{The relative importance of the interference terms
in the \ZHtrans\ and helicity bases 
at $\protect\sqrt{s}=250\GeV$ and $m_h = 120\GeV$ ($\bz = 0.59$)
for (A)~$\eebar\rightarrow Z\HIGGS \rightarrow b\bar{b}\HIGGS$
and (B)~$\eebar\rightarrow ZA \rightarrow b\bar{b}A$.
Plotted is the differential
distribution in $\Ihat$, the value of the interference term
normalized to the square of the total matrix element.
}
\label{ZHZAinf}
\end{figure*}
comparing the distribution in $\Ihat$ for the unpolarized processes
$e^{+} e^{-} \longrightarrow Z\HIGGS \longrightarrow b\bar{b}\HIGGS$ 
and $e^{+}e^{-}\longrightarrow ZA \longrightarrow b\bar{b}A$
using the \ZHtrans\ and
helicity bases, and assuming that $M_\HIGGS = 120$~GeV and
$\sqrt{s} = 250$~GeV ($\bz = 0.59$).
Fig.~\ref{ZHZAinf}A shows that the removal from
$Z\HIGGS$ production of the longitudinal $Z$'s with their
relatively wide distribution in $\cos\chi$ leads to a greatly
reduced role for the interference terms in the \ZHtrans\ basis
as compared to the helicity basis.
In fact, nearly 85\% of the total cross section is accounted
for by the $\vert\Ihat\vert < 0.10$ region when using the
\ZHtrans\ basis: this is a reasonably narrow distribution.
On the other hand, since all three spin states
contain significant populations in the helicity basis, the interference
terms tend to be large a significant fraction of the time;
in fact, only 31\% of the cross section comes from configurations where 
$\vert\Ihat\vert < 0.10$.
In Table~\ref{Zh-InfInteg} we have tabulated the fraction of the
total cross section contributed by regions where 
$\vert\Ihat\vert<0.05$,
0.10, 0.15, 0.20, and 0.25 in these two bases.
\begin{table*} 
\caption{Fraction of the 
$\eebar\rightarrow Z\HIGGS \rightarrow b\bar{b} \HIGGS$ cross
section coming from phase space points where the interference
term is less than or equal to some cutoff
assuming a Higgs mass of 120~GeV
and a machine center-of-mass energy of 250~GeV ($\bz = 0.59$).  
\label{Zh-InfInteg}}
\begin{ruledtabular}
\begin{tabular}{ccccc}
&Integration   & \hbox{Helicity} &  \hbox{\ZHtrans} &  \\
&region        & \hbox{basis}    &  \hbox{basis}  &  \\[0.05in]
\hline
&$\vert\widehat{\cal I}\vert < 0.05$ &  0.159   &    0.640 &  \\    
&$\vert\widehat{\cal I}\vert < 0.10$ &  0.311   &    0.844 &  \\    
&$\vert\widehat{\cal I}\vert < 0.15$ &  0.454   &    0.950 &  \\    
&$\vert\widehat{\cal I}\vert < 0.20$ &  0.595   &    0.987 &  \\    
&$\vert\widehat{\cal I}\vert < 0.25$ &  0.734   &    0.998 &  \\    
\end{tabular}
\end{ruledtabular}
\end{table*}
\begin{table*} 
\caption{Fraction of the 
$\eebar\rightarrow ZA\rightarrow b\bar{b} A$ cross
section coming from phase space points where the interference
term is less than or equal to some cutoff
assuming a Higgs mass of 120~GeV
and a machine center-of-mass energy of 250~GeV ($\bz = 0.59$).  
\label{ZA-InfInteg}}
\begin{ruledtabular}
\begin{tabular}{ccccc}
&Integration   & \hbox{Helicity} &  \hbox{\ZHtrans} &  \\
&region        & \hbox{basis}    &  \hbox{basis}  &  \\[0.05in]
\hline
&$\vert\widehat{\cal I}\vert < 0.05$ &  0.406   &    0.291 &  \\    
&$\vert\widehat{\cal I}\vert < 0.10$ &  0.561   &    0.461 &  \\    
&$\vert\widehat{\cal I}\vert < 0.15$ &  0.662   &    0.589 &  \\    
&$\vert\widehat{\cal I}\vert < 0.20$ &  0.737   &    0.695 &  \\    
&$\vert\widehat{\cal I}\vert < 0.25$ &  0.797   &    0.785 &  \\    
\end{tabular}
\end{ruledtabular}
\end{table*}

Fig.~\ref{ZHZAinf}B displays the same information for the
$e^{+}e^{-}\longrightarrow ZA \longrightarrow b\bar{b}A$
case.   
In this situation, the contrast between the
``right'' (helicity) and ``wrong'' (\ZHtrans) bases is less dramatic:
both choices of spin basis contain significant contributions from
regions where the interference terms are large:  in particular,
both distributions exhibit long tails in the $\Ihat<0$ (destructive
interference) region.  From Table~\ref{ZA-InfInteg}
we learn that the range $\vert\Ihat\vert < 0.10$ contributes
56\% of the total cross section in the helicity basis but only
46\% in the \ZHtrans\ basis.

\vfill\eject\section{Lorentz-covariant form} 
Even when we include
the decay of the $Z$ boson, the Lorentz-invariant result for the
square of the matrix element for associated Higgs production
is surprisingly simple.
If we let each particle's 4-momentum be represented by
its symbol ($e \equiv e^{-}$ and $\bar{e} \equiv e^{+}$),
then we obtain
\beqa
\Bigl\vert{\cal M}(e_L^{-}e_R^{+} \rightarrow Z\HIGGS
\rightarrow f \bar{f} \HIGGS 
)\Bigr\vert^2 &\sim&
4q_{eL}^2 \Biggl[ q_{fL}^2(2\bar{e}\cdot{f})(2e\cdot\bar{f})
+ q_{fR}^2(2e\cdot{f})(2\bar{e}\cdot\bar{f}) \Biggr]
\label{WZ770}
\eeqa
and
\beqa
\Bigl\vert{\cal M}(e_R^{-}e_L^{+} \rightarrow Z\HIGGS
\rightarrow f \bar{f} \HIGGS 
)\Bigr\vert^2 &\sim&
4q_{eR}^2 \Biggl[ q_{fL}^2(2e\cdot{f})(2\bar{e}\cdot\bar{f})
+ q_{fR}^2(2\bar{e}\cdot{f})(2e\cdot\bar{f})\Biggr].
\label{WZ771}
\eeqa
for a $CP$-even Higgs.  The results for a $CP$-odd Higgs
are not quite as compact as Eqs.~(\ref{WZ770}) and~(\ref{WZ771}),
primarily because of the momentum-dependent Levi-Cevita-tensor
vertex:
\beqa
\Bigl\vert{\cal M}(e_L^{-}e_R^{+} \rightarrow ZA
\rightarrow f \bar{f} A 
)\Bigr\vert^2 &\sim&
q_{eL}^2 q_{fL}^2 \Biggl\{ (2{e}\cdot\bar{e}) (2{f}\cdot\bar{f})
                  \Bigl[ (2{f}\cdot\bar{e})^2 
                        +(2{e}\cdot\bar{f})^2
\cr && \qquad\qquad\qquad\qquad\qquad
                        +2(2{e}\cdot{f})(2\bar{e}\cdot\bar{f})
                        -(2{e}\cdot\bar{e})(2{f}\cdot\bar{f})
                  \Bigr] 
\cr && \qquad\qquad
- \Bigl[ (2{e}\cdot\bar{f})(2{f}\cdot\bar{e})
                                 -(2{e}\cdot{f})(2\bar{e}\cdot\bar{f})
                           \Bigr]^2
                  \Biggr\}
\cr &&
+q_{eL}^2 q_{fR}^2 \Biggl\{ (2{e}\cdot\bar{e}) (2{f}\cdot\bar{f})
                  \Bigl[ (2{e}\cdot{f})^2 
                        +(2\bar{e}\cdot\bar{f})^2
\cr && \qquad\qquad\qquad\qquad\qquad
                        +2(2\bar{e}\cdot{f})(2{e}\cdot\bar{f})
                        -(2{e}\cdot\bar{e})(2{f}\cdot\bar{f})
                  \Bigr] 
\cr && \qquad\qquad
- \Bigl[ (2{e}\cdot{f})(2\bar{e}\cdot\bar{f})
                                 -(2{e}\cdot\bar{f})(2\bar{e}\cdot{f})
                           \Bigr]^2
                  \Biggr\}
\label{WZ773-L}
\eeqa
\beqa
\Bigl\vert{\cal M}(e_R^{-}e_L^{+} \rightarrow ZA
\rightarrow f \bar{f} A) 
\Bigr\vert^2 &\sim&
q_{eR}^2 q_{fL}^2 \Biggl\{ (2{e}\cdot\bar{e}) (2{f}\cdot\bar{f})
                  \Bigl[ (2{e}\cdot{f})^2 
                        +(2\bar{e}\cdot\bar{f})^2
\cr && \qquad\qquad\qquad\qquad\qquad
                        +2(2\bar{e}\cdot{f})(2{e}\cdot\bar{f})
                        -(2{e}\cdot\bar{e})(2{f}\cdot\bar{f})
                  \Bigr] 
\cr && \qquad\qquad
- \Bigl[ (2{e}\cdot{f})(2\bar{e}\cdot\bar{f})
                                 -(2{e}\cdot\bar{f})(2\bar{e}\cdot{f})
                           \Bigr]^2
                  \Biggr\}
\cr &&
+q_{eR}^2 q_{fR}^2\Biggl\{ (2{e}\cdot\bar{e}) (2{f}\cdot\bar{f})
                  \Bigl[ (2{f}\cdot\bar{e})^2 
                        +(2{e}\cdot\bar{f})^2
\cr && \qquad\qquad\qquad\qquad\qquad
                        +2(2{e}\cdot{f})(2\bar{e}\cdot\bar{f})
                        -(2{e}\cdot\bar{e})(2{f}\cdot\bar{f})
                  \Bigr] 
\cr && \qquad\qquad
- \Bigl[ (2{e}\cdot\bar{f})(2{f}\cdot\bar{e})
                                 -(2{e}\cdot{f})(2\bar{e}\cdot\bar{f})
                           \Bigr]^2
                  \Biggr\}.
\label{WZ773-R}
\eeqa

\vfill\eject\section{The Beamline Basis}\label{BML}

Another potentially interesting basis is the 
beamline basis~\cite{ttbar1}, which is defined by
\beq
\sin\xi = { { \ginv\sin\thetas }  \over { 1-\bz\cos\thetas} };
\qquad
\cos\xi = { {\cos\thetas - \bz} \over { 1-\bz\cos\thetas} }.
\label{BMLdef}
\eeq
Here $\bz$ is the ZMF speed of the $Z$ boson.
In this basis, the spin axis for the $Z$ is the electron direction.
Because the beam directions are experimentally well-determined, 
any angular measurement uncertainty issues
in this basis should be very different than those in
the helicity or \ZHtrans\ bases.

Figure~\ref{CTHCXI} displays the relationship between $\cos\xi$
and $\cos\theta^*$ in the beamline, \ZHtrans\ and helicity bases
for low ($\BETA=0.11$), medium ($\BETA=0.59$) and high
($\BETA=0.92$) values of the ZMF speed of the $Z$ boson.
Near threshold ($\BETA\rightarrow 0$), the beamline basis is
nearly coincident with the \ZHtrans\ basis.  In the other extreme
($\BETA\rightarrow 1$), the beamline basis becomes coincident
with the helicity basis.  Between these two extremes, the beamline
basis interpolates between these two bases.
\begin{figure*}
\vspace*{15cm}
\includegraphics{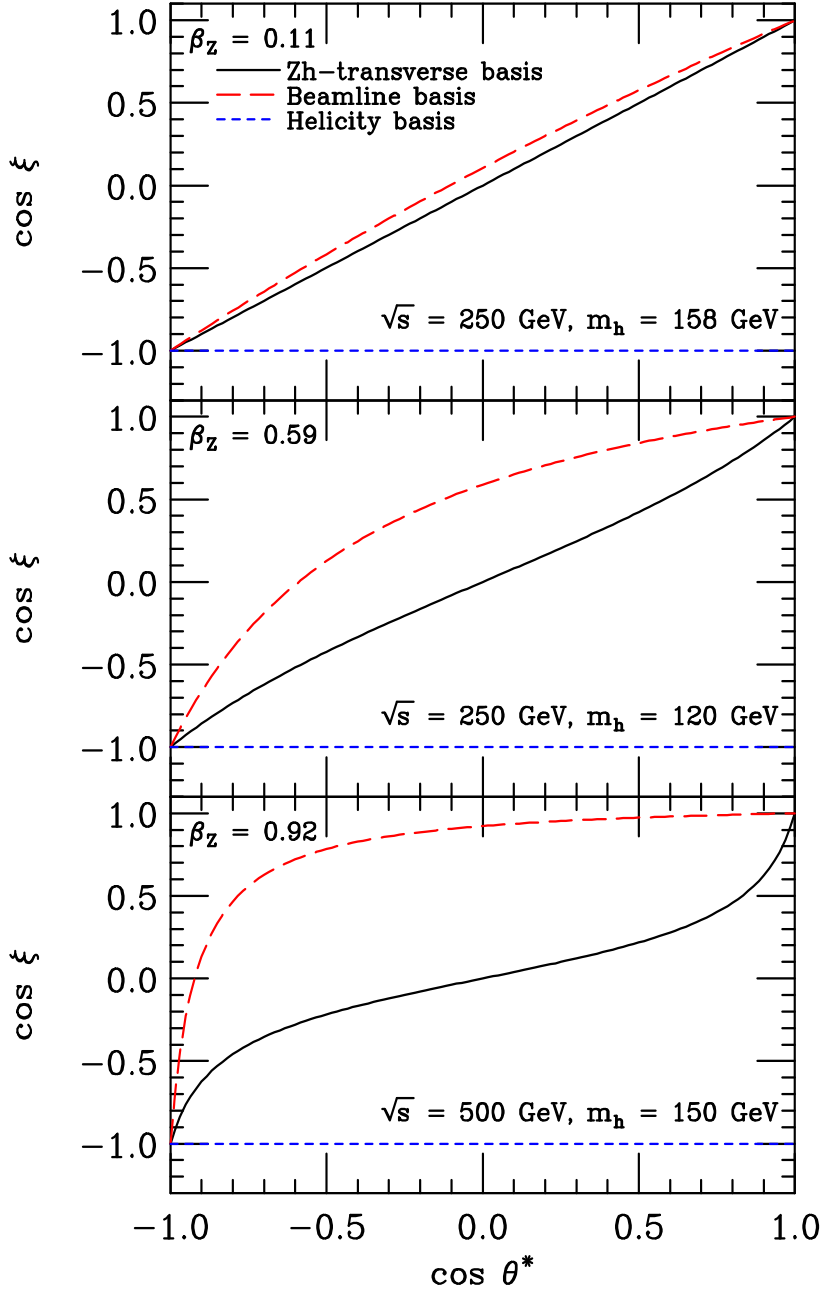}
\vspace{4.0cm}
\caption[]{The relationship between $\cos\xi$ and $\cos\theta^*$
in the \ZHtrans\ (solid), beamline (long dashes), and helicity 
(short dashes) bases at 
$\BETA = 0.11$, $0.59$, and $0.92$.
At small $\BETA$, the beamline and \ZHtrans\ bases are approximately
the same.  For $\BETA\rightarrow1$, the beamline basis becomes the
$\cxi=1$ variant of the 
helicity basis while the \ZHtrans\ approaches
the basis with $\cxi \equiv 0$.
}
\label{CTHCXI}
\end{figure*}

Figure~\ref{CTH-BML} illustrates the breakdown of the $Z\HIGGS$
and $ZA$ production cross sections at $\BETA = 0.59$
into the various $Z$ spin
states as a function of the production angle in the ZMF for
both possible polarizations of the colliding beams.  Numerical
values for these fractions for a Higgs mass of 120 GeV and
a collider center-of-mass energy of 250 GeV
$\sqrt{s} = 250$~GeV are presented in Table~\ref{BMLfractions}.
\begin{table*}
\caption{Spin decompositions in the beamline basis for 
$\eebar \rightarrow Z\phi$
assuming $m_\phi = 120 \GeV$ and $\sqrt{s} = 250 \GeV$
($\beta_{{}_Z}=0.59$).
\label{BMLfractions}}
\begin{ruledtabular}
\begin{tabular}{crrr}
 & $(+)$ &  $(-)$ &  $(0)$  \\[0.05in]
\hline
$Z\HIGGS$ &&& \\
\hline
$e_L^- e_R^+$           & 84.8\% & 0.0\% &  15.2\%  \\
$e_R^- e_L^+$           & 0.0\% & 84.8\% &  15.2\%  \\
$e^- e^+$ (unpolarized) & 48.6\% & 36.2\% & 15.2\% \\
\hline
$ZA$ &&& \\
\hline
$e_L^- e_R^+$            & 50.0\% &  8.3\% &  41.7\% \\
$e_R^- e_L^+$            &  8.3\% & 50.0\% &  41.7\% \\
$e^- e^+$ (unpolarized)  & 32.2\% & 26.1\% &  41.7\% \\
\end{tabular}
\end{ruledtabular}
\end{table*}
Recall that
the helicity basis is optimal for pseudoscalar production and
that, as shown on Fig.~\ref{CTHCXI}, the beamline basis is far
from the helicity basis at this energy.  Thus, it is not surprising
to find significant contributions from all three spin states
to the $ZA$ cross section.  On the other hand, for $Z\HIGGS$
production, the beamline basis does zero out one of the transverse
spin states when polarized beams are used.
\begin{figure*}
\vspace*{13cm}
\includegraphics{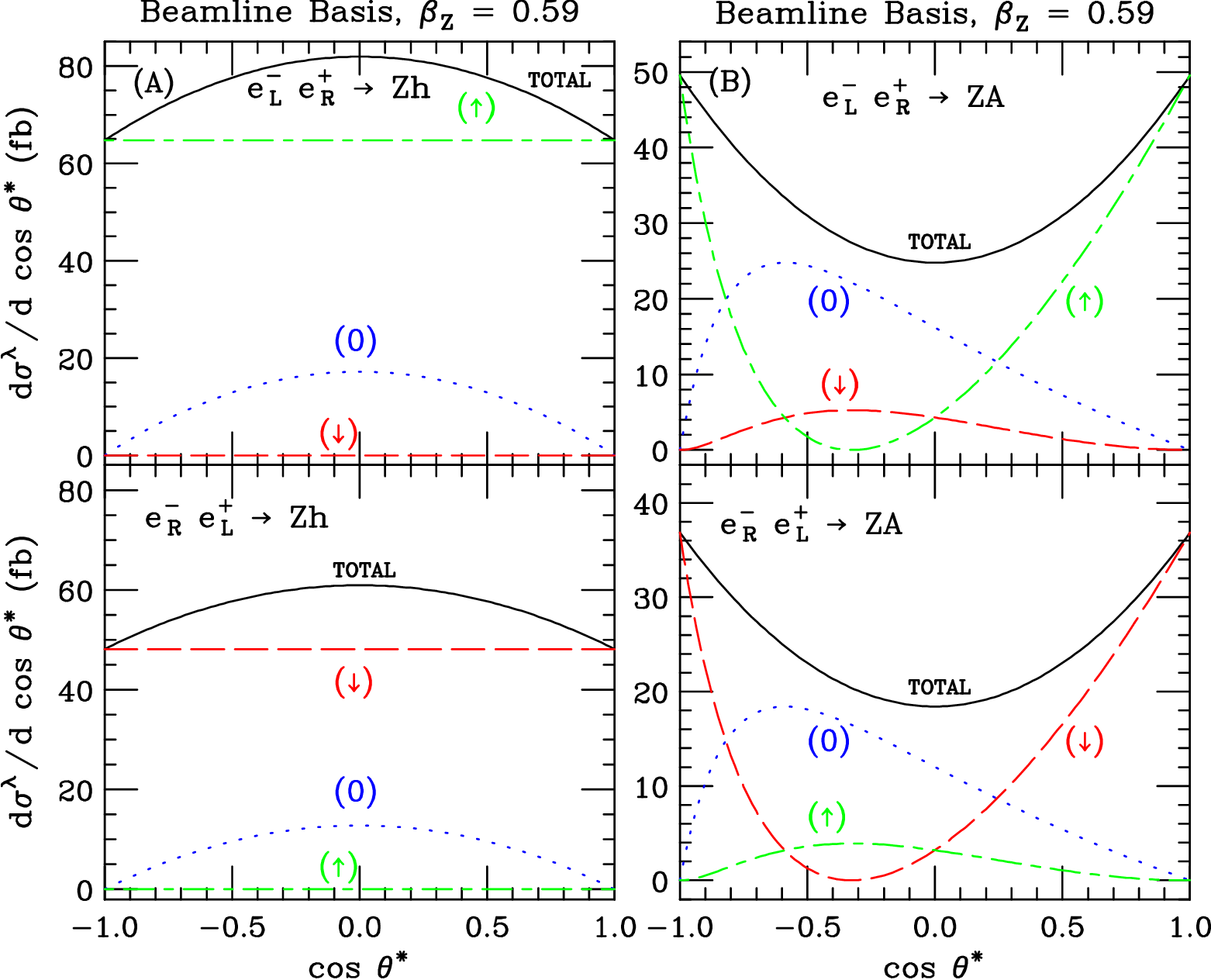}
\caption[]{Polarized production angle distributions for 
{\bf (A)}\ $\eebar\rightarrow Z\HIGGS$ 
and
{\bf (B)}\ $\eebar\rightarrow ZA$
assuming $\sqrt{s} = 250$ GeV and $M_h = 120$ GeV
($\beta_{{}_{Z}}=0.59$).  
Displayed are 
the contributions from the three possible $Z$ spins in the
beamline basis.
}
\label{CTH-BML}
\end{figure*}

Fig.~\ref{beta-BML}
illustrates the evolution of the spin fractions 
in the beamline basis as the machine energy is changed.
Amusingly, one of the two transverse spin components is always
equal to  50\% in  this basis for $ZA$ production, with the remainder
divided between the other transverse spin and the longitudinal
component.  

The explicit form of the spin functions in the beamline basis read
\beqa
\T_L^{+} &=& \GAMMA^{-1} \sqrt{2}  \cr
\T_L^{-} &=& 0                     \cr
\T_L^{0} &=& \BETA \sth
\eeqa
for $Z\HIGGS$ production, and 
\beqa
\widetilde\T_L^{+} &=&  {{1}\over{\sqrt{2}}}\ts
{{2\cth-\BETA(1+\cth^2)}\over{1-\BETA\cth}}     \cr
\widetilde\T_L^{-} &=&  {{1}\over{\sqrt{2}}}\ts
{{\BETA\sth^2}\over{1-\BETA\cth}}     \cr
\widetilde\T_L^{0} &=& {{\GAMMA^{-1}\sth}\over{1-\BETA\cth}}
\eeqa
for $ZA$ production.
\begin{figure*}
\vspace*{13cm}
\includegraphics{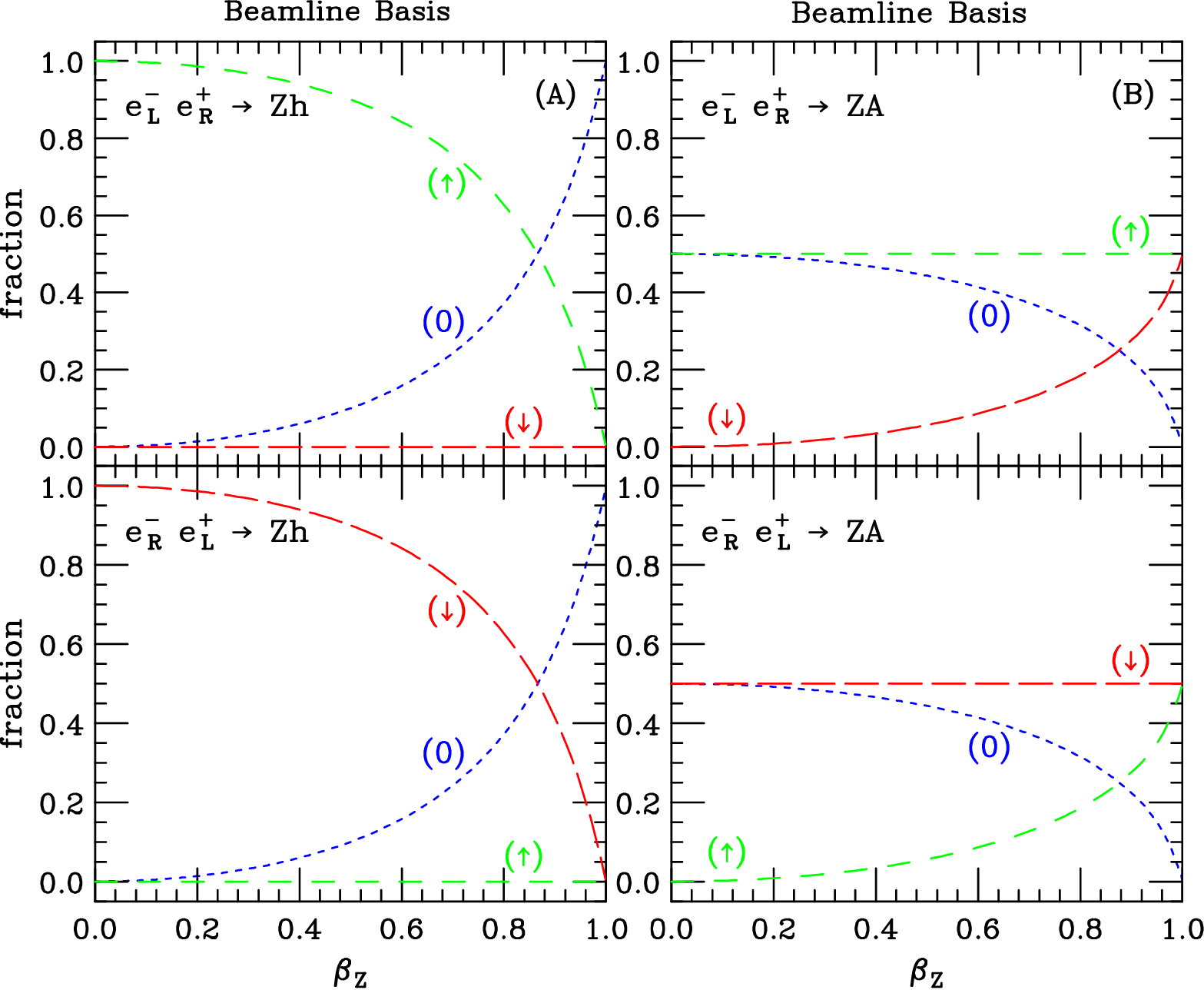}
\caption[]{Spin decomposition in the beamline basis
of the polarized associated Higgs production
cross sections as a function of the ZMF speed $\bz$ of the $Z$
boson. 
Shown are the fractions of
the total cross sections in the $(\uparrow)$, $(\downarrow)$,
and $(0)$ spin states in {\bf (A)} $Z\HIGGS$ and {\bf (B)} $ZA$
production.
The beamline basis interpolates between the $Z\HIGGS$-transverse
basis at low $\BETA$ and the helicity basis at high $\BETA$
as can be seen by comparison with Figs.~\protect\ref{ZHbetaplot}
and~\protect\ref{ZAbetaplots}.
}
\label{beta-BML}
\end{figure*}

Turning to the $Z$ decay angular distributions discussed in
Sec.~\ref{enchiladas}, we obtain the following expressions
in the beamline basis:  first, for $e_{L}^{-} e_{R}^{+} \rightarrow
Z\HIGGS$
\beqa
{{1}\over{\sigma_{L}}}\ts
{{d\sigma_L}\over{d(\cos\chi)}} 
\Biggr\vert_{\rm beamline} &&
= {{9}\over{8}} {{1-\BETA^2}\over{3-2\BETA^2}} (1 + \cos^2\chi)
+ {{3}\over{4}} {{\BETA^2}\over{3-2\BETA^2}} \sin^2\chi
\cr && \quad + {{9}\over{4}} 
{{1-\BETA^2}\over{3-2\BETA^2}} 
{
{ q_{fL}^2 - q_{fR}^2 }
\over
{ q_{fL}^2 + q_{fR}^2 }
} \ts
\cos\chi,
\label{WZ692B}
\eeqa
and for $e_{L}^{-} e_R^+ \rightarrow ZA$:

\noindent
\beqa
{{1}\over{\sigma_{L}}}\ts
{{d\sigma_L}\over{d(\cos\chi)}} 
\Biggr\vert_{\rm beamline} &&
= 
\left\{ 
         {{3}\over{32}} 
        +{{9}\over{32}}{{1}\over{\BETA^2}}
         \Biggl[   2 - \BETA^2 
                   + {{\GAMMA^{-2}}\over{\BETA}} 
                     \ln \Bigl( {{1-\BETA}\over{1+\BETA}} 
                         \Bigr)
         \Biggr]
\right\}
(1 + \cos^2\chi)
\cr && \quad 
- {{9}\over{16}} {{\GAMMA^{-2}}\over{\BETA^2}} 
\Biggl[ 2 + {{1}\over{\BETA}} 
\ln \Bigl( {{1-\BETA}\over{1+\BETA}} \Bigr)
\Biggr]
\sin^2\chi
\cr && \quad - {{9}\over{16}}\ts
{
{ q_{fL}^2 - q_{fR}^2 }
\over
{ q_{fL}^2 + q_{fR}^2 }
} \ts
{{\GAMMA^{-2}}\over{\BETA^2}} 
\Biggl[ 2 + {{1}\over{\BETA}} 
\ln \Bigl( {{1-\BETA}\over{1+\BETA}} \Bigr)
\Biggr]
\ts
\cos\chi,
\label{WZ743Z}
\eeqa
These distributions are displayed in Fig.~\ref{enchilada-BML}
for both the scalar and pseudoscalar Higgs cases.  
For moderate $\BETA$, the separation
between $Zh$ and $ZA$ is nearly as good as in the \ZHtrans\ basis.
The values of the $\cos\chi$ forward-backward asymmetry ratio,
Eq.~(\ref{EnchiladaRatio}),
in the beamline basis are
\beq
\ECKS_{\rm beamline}^{Zh} = 3.96; \qquad
\ECKS_{\rm beamline}^{ZA} = 1.83.
\eeq
Comparison to Eq.~(\ref{ECKS-transverse}) reveals that
in terms of this particular measure, the beamlinie basis 
performs nearly as well as the $Z\HIGGS$-transverse basis.
\begin{figure*}
\vspace*{8cm}
\includegraphics{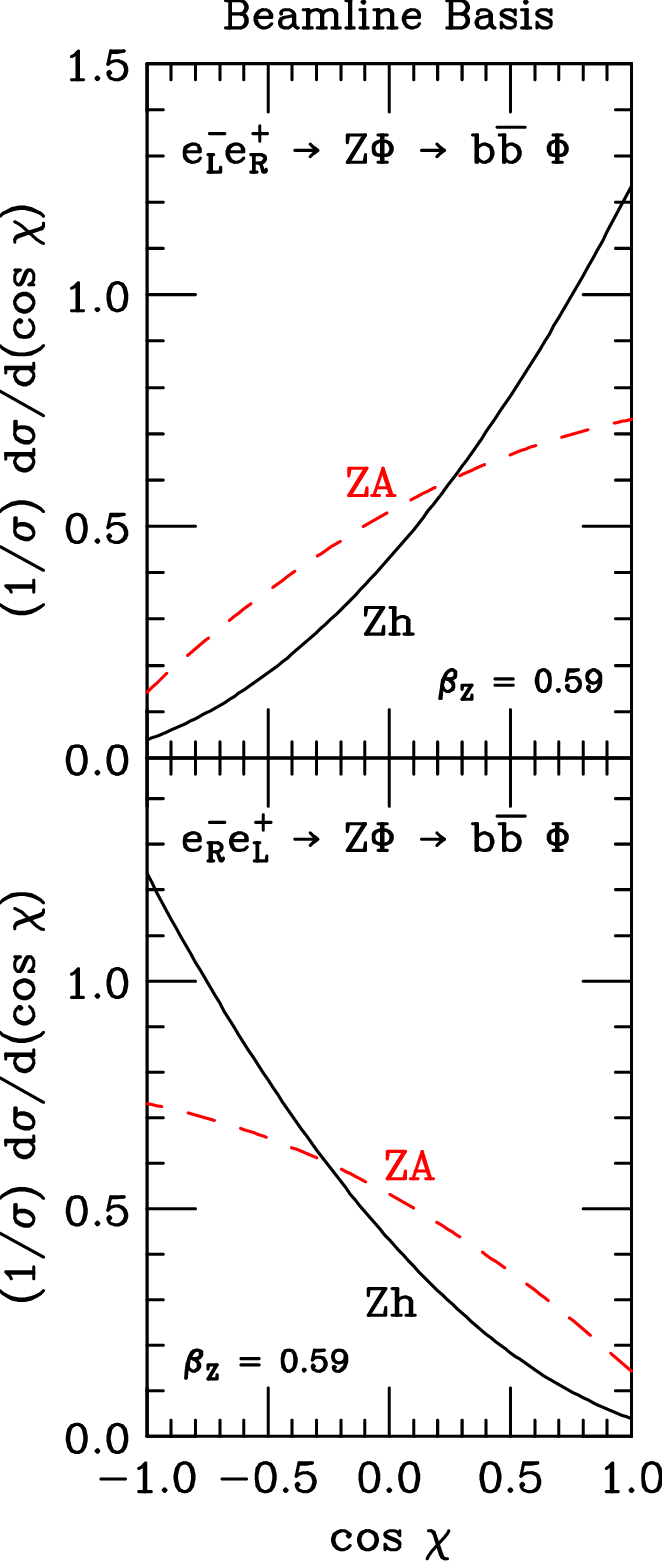}
\vspace{8.5cm}
\caption[]{Comparison of $Z$ decay angular distributions for
$\eebar\rightarrow Zh\rightarrow b \bar{b} h$ 
and
$\eebar\rightarrow ZA\rightarrow b \bar{b}A$ 
in the beamline basis
for $\BETA=0.59$ ($M_\HIGGS = 120 \GeV$, $\protect\sqrt{s} = 250 \GeV$).
The decay angle $\chi_{{}_Z}$ is
defined as the angle in the $Z$ rest frame between 
the spin axis direction and the direction of motion of the 
negatively-charged lepton.
}
\label{enchilada-BML}
\end{figure*}




\begin{references}
\bibitem{assume}
Even if we don't assume
the Standard Model coupling on the $ee\HIGGS$ vertex, the absence of
a $e^{+}e^{-}\longrightarrow \HIGGS$ signal at LEP ensures that this
diagram is negligible~\protect\cite{Miller}.
\bibitem{Lambda} For the
MSSM, $\Lambda$ would be related to the SUSY-breaking scale and
$\eta$ would depend on the particle content appearing in the loops.
\bibitem{messy} Inclusion
of the finite mass effects would result in straightforward but
messy modifications to Eqs.~(\protect\ref{DecayFactors-L}) 
and~(\protect\ref{DecayFactors-R})   These effects are greatest for
$b$-quarks, where they are less than a few percent.
\end{references}
\end{document}